\documentclass[final,onecolumn,prf,aps,amsfonts,amssymb,showpacs,superscriptaddress,floatfix,nofootinbib]{revtex4-1}
\usepackage{amsmath}
\usepackage{graphicx}
\usepackage[dvipsnames]{color}
\usepackage{hyperref}
\usepackage{psfrag}
\usepackage{stmaryrd}
\usepackage{amssymb}
\usepackage{wasysym}
\usepackage{float}
\usepackage{color}

\usepackage[latin1]{inputenc}
\usepackage[T1]{fontenc}
\usepackage[french,english]{babel}

\usepackage{geometry}
\DeclareMathOperator{\sign}{sign}
\DeclareRobustCommand{\Re}{\operatorname{\mathfrak{Re}}}
\DeclareRobustCommand{\Im}{\operatorname{\mathfrak{Im}}}

\begin{document}

\title{Wake of inertial waves of a horizontal cylinder in horizontal translation}

\author{Nathana\"el~Machicoane}
\affiliation{Laboratoire FAST, CNRS, Universit\'e Paris-Sud,
Universit\'e Paris-Saclay, 91405 Orsay, France}
\author{Vincent~Labarre}
\affiliation{Laboratoire FAST, CNRS, Universit\'e Paris-Sud,
Universit\'e Paris-Saclay, 91405 Orsay, France}
\author{Bruno~Voisin}
\affiliation{Laboratoire LEGI, CNRS, Grenoble INP, Universit\'e
Grenoble Alpes, 38058 Grenoble, France}
\author{Frédéric~Moisy}
\affiliation{Laboratoire FAST, CNRS, Universit\'e Paris-Sud,
Universit\'e Paris-Saclay, 91405 Orsay, France}
\author{Pierre-Philippe~Cortet}
\email[]{ppcortet@fast.u-psud.fr} \affiliation{Laboratoire FAST,
CNRS, Universit\'e Paris-Sud, Universit\'e Paris-Saclay, 91405
Orsay, France}

\date{\today}

\begin{abstract}
We analyze theoretically and experimentally the wake behind a
horizontal cylinder of diameter $d$ horizontally translated at
constant velocity $U$ in a fluid rotating about the vertical axis
at a rate $\Omega$. Using particle image velocimetry measurements
in the rotating frame, we show that the wake is stabilized by
rotation for Reynolds number ${\rm Re}=Ud/\nu$ much larger than in
a non-rotating fluid. Over the explored range of parameters, the
limit of stability is ${\rm Re} \simeq (275 \pm 25) / {\rm Ro}$,
with ${\rm Ro}=U/2\Omega d$ the Rossby number, indicating that the
stabilizing process is governed by the Ekman pumping in the
boundary layer. At low Rossby number, the wake takes the form of a
stationary pattern of inertial waves, similar to the wake of
surface gravity waves behind a ship. We compare this steady wake
pattern to a model, originally developed by [Johnson, J.\ Fluid
Mech.\ \textbf{120}, 359 (1982)], assuming a free-slip boundary
condition and a weak streamwise perturbation. Our measurements
show a quantitative agreement with this model for ${\rm
Ro}\lesssim 0.3$. At larger Rossby number, the phase pattern of
the wake is close to the prediction for an infinitely small line
object. However, the wake amplitude and phase origin are not
correctly described by the weak-streamwise-perturbation model,
calling for an alternative model for the boundary condition at
moderate rotation rate.
\end{abstract}

\maketitle

\section{Introduction}

Since the celebrated work of Taylor~\cite{Taylor1923} in 1923, it
is known that a solid object in slow horizontal motion in a fluid
rapidly rotating about the vertical axis tends to drive with it a
vertical column of fluid circumscribing the object, as if they
were forming together a solid body. The fluid outside of this
``Taylor column'' flows around it, remaining in the same
horizontal plane. This is consistent with the Taylor--Proudman
theorem, which states that linear and inviscid fluid motions
associated to timescales much longer than the global rotation
period must be vertically invariant~\cite{GreenspanBook}. Another
major feature of rotating fluids is their ability to propagate a
specific class of waves, called inertial waves, which are both
anisotropic and dispersive~\cite{GreenspanBook}. Here, we study
the flow around a 2D horizontally invariant cylinder in horizontal
translation. This geometry is of interest because the ``Taylor
column'' solution is prevented by mass conservation  even in the
limit of large rotation rate (the fluid cannot flow around the
cylinder), so only the inertial wave solution is expected for the
wake. The aim of this paper is to compare velocity measurements in
this configuration with a theoretical approach based on an
approximation of weak streamwise perturbation, originally proposed
by Johnson~\cite{Johnson1982}. Although this approximation may
seem unnatural for non-slender bodies, it can be shown to apply in
the limit of strong background rotation, and is indeed found to
compare well with our measured wake patterns in this regime.

The general problem of an object moving horizontally in a fluid
under rotation has received a lot of
attention~\cite{Hide1966,Lighthill1967,Hide1968,Redekopp1975,Peat1976,Stewartson1979,Mason1981,Johnson1982,Cheng1982,Heikes1982}.
Several flow regimes have been reported depending on the relative
importance of rotation, non-linearities and viscous effects,
characterized by the Rossby number ${\rm Ro}=U/2\Omega L$ and the
Reynolds number ${\rm Re}=UL/\nu$, but also on the object height
relative to the fluid depth $h/H$ and the object aspect ratio
$h/L$ ($\Omega$ is the rotation rate, $U$ the object velocity, and
$L$ its size in the streamwise direction). In the strong rotation
limit ${\rm Ro} \ll 1$, the flow can be of two kinds depending on
the geometry: either (i) a geostrophic (nearly) vertically
invariant horizontal flow, i.e.\ the Taylor-column
flow~\cite{Hide1966,Hide1968,Stewartson1979,Mason1981}, which
cannot be decomposed in terms of inertial waves, or (ii) a wake of
inertial
waves~\cite{Lighthill1967,Hide1968,Redekopp1975,Peat1976,Stewartson1979,Mason1981,Johnson1982,Cheng1982,Heikes1982}.
This second kind, akin to the wake of gravity surface waves behind
a ship~\cite{Lighthill1978,Darrigol2005}, is expected when the
fluid flows, at least partially, over and/or below the object,
locally inducing a vertical velocity perturbation which triggers
the emission of inertial waves. The phase configuration of such
wake of inertial waves has been derived theoretically for
infinitely small objects by Lighthill~\cite{Lighthill1967} and
Redekopp~\cite{Redekopp1975} and further explored experimentally
for a sphere by Hide, Ibbetson and Lighthill~\cite{Hide1968} (dye
observations) and for a cross-stream cylinder by Peat and
Stevenson~\cite{Peat1976} (schlieren observations). In these
works, the question of which region of the wake pattern is
supplied with energy was however not addressed: to do so, it is
necessary to model the wave-field boundary condition close the
object.

The duality between the low-$\rm Ro$ Taylor column solution and
the finite-$\rm Ro$ wake of inertial waves has remained unclear
for a long time. An important step was made by Hide and
Ibbetson~\cite{Hide1966} who predicted and verified experimentally
that, for objects of comparable vertical $h$ and horizontal $L$
sizes, the Taylor column appears when the Rossby number $\rm Ro$
becomes lower than typically $h/H$ (see also Mason and
Sykes~\cite{Mason1981}). Hide and Ibbetson also exhibited
experimentally a reduction of this $\rm Ro$-threshold when the
Ekman number ${\rm Ek}={\rm Ro}/{\rm Re}$ becomes larger than
$\sim 10^{-3}$ due to the growing role of viscous boundary layers.

A more precise solution to this duality between Taylor column and
wake of inertial waves was proposed for slender bodies ($h/L\ll
1$) in 1979 by Stewartson and Cheng~\cite{Stewartson1979} who
demonstrated the bimodal nature of the flow. They predicted
theoretically the inviscid flow produced by the horizontal
translation of a thin object by taking the limit ${\rm Ro}\ll 1$
at fixed parameter $H\,{\rm Ro}/L$, and linearizing the inviscid
boundary condition using the slender body assumption. Their
analysis, though not employing the term, is quasi-geostrophic,
replacing the horizontal velocity by its geostrophic value in the
acceleration term of the horizontal momentum equation, therefore
implying long time scale and large vertical scale. For an object
of comparable horizontal streamwise and cross-stream lengths $L$,
the Taylor-column flow is dominant when the parameter $H\,{\rm
Ro}/L$ is small, whereas the wake of inertial waves dominates when
it is large.

Johnson~\cite{Johnson1982} and Cheng and Johnson~\cite{Cheng1982}
extended the description of Stewartson and
Cheng~\cite{Stewartson1979} to a viscous fluid of arbitrary depth.
Johnson~\cite{Johnson1982} studied in particular the case of 2D
objects, invariant along the horizontal cross-stream direction,
and showed that in this geometry the inviscid non-penetration
boundary condition can be simplified even in the case of
non-slender objects ($h/L \sim 1$) with the assumption of weak
streamwise perturbation. In this geometry, the cross-stream
invariance also prevents the emergence of a Taylor-column flow and
therefore leads to a pure wake of inertial waves down to vanishing
Rossby number. The analysis of Johnson also shows how the shape
and finite size of the 2D object modify the wake pattern compared
to that of an infinitely small source.

Heikes and Maxworthy~\cite{Heikes1982} tested experimentally the
predictions of Johnson~\cite{Johnson1982} by studying the
perturbation of a horizontal flow by a ridge made of a portion of
a cylinder. They used aluminium flakes to highlight shear regions
but also to draw flow streamlines on long exposure images. Their
observations revealed an upstream shift of the oscillations
compared to the theory as well as a wake amplitude smaller than
its theoretical prediction. Heikes and Maxworthy did however not
consider viscous dissipation in their model and the reasons
(non-linear effects, viscous effects) for the discrepancies of
their experiments with theory remained unclear.

Since then, the range of validity of the
weak-streamwise-perturbation approximation for a 2D non-slender
object has remained an open question. In this article, we provide
a quantitative test of this theory by measuring the wake of a
horizontal cylinder in a rotating water-filled tank using particle
image velocimetry. Our theoretical approach retains the
weak-streamwise-perturbation and infinite-depth approximations of
Johnson~\cite{Johnson1982}, but relaxes the quasi-geostrophic
approximation, shown by Heikes and Maxworthy~\cite{Heikes1982} to
hold for small Rossby number only (typically below $10^{-1}$).
Viscosity effects are considered in the bulk only, while an
inviscid free-slip boundary condition is kept along the object. We
also address the stability of the wake in terms of the control
parameters (Reynolds and Rossby numbers). This problem, which has
not been addressed yet to the best of our knowledge, is related to
the question of the separation and stability of boundary layers on
non-vertical surfaces in a fluid rotating about the vertical axis.

After recalling the derivation of the phase field of the inviscid
wake of inertial waves of a line object in Sec.~\ref{sec:line}, we
derive in Sec.~\ref{sec:theory} the velocity field of the steady
wake of a cylinder of diameter $d$ using the
weak-streamwise-perturbation approximation of
Johnson~\cite{Johnson1982}. In the experimental
section~\ref{sec:expresults}, we first study the threshold in
Reynolds and Rossby numbers above which the wake becomes unsteady.
These data reveals the strong stabilization of the wake by
rotation (Sec.~\ref{sec:unsteady}): we show that the wake remains
steady up to ${\rm Re}\sim 1000$ at ${\rm Ro}\sim 0.3$, a value
much larger than the onset of the K\'{a}rm\'{a}n vortex street in
a non-rotating fluid. We finally show in Sec.~\ref{sec:steady}
that the theory of Sec.~\ref{sec:theory} describes quantitatively
the steady wake of inertial waves for ${\rm Ro} \lesssim 0.3$ and
for ${\rm Re}$ ranging from order $1$ to $10^3$. These
measurements show that the inviscid boundary condition associated
to the weak-streamwise-perturbation approximation considered here
is valid even for a non-slender 2D object. At larger Rossby
numbers, for which the finite size of the object no longer
determines the structure of the wave field, we recover
experimentally the wake predicted for a line object. However, we
show that the weak-streamwise-perturbation approximation does not
describe correctly the amplitude and phase origin of the wake.
These last observations call for a better understanding of the
nature of the boundary layers on the object at moderate rotation
rates.

\section{Linear wake of a 2D object in horizontal translation in a rotating
fluid}\label{sec:theo}

We describe here the wake produced by the translation at constant
velocity ${\bf U}=U{\bf e}_x$ of a 2D object, invariant along
${\bf e}_y$, in a fluid rotating at rate $\Omega$ about ${\bf
e}_z$.  Small perturbations satisfy the linearized Navier--Stokes
equation
\begin{equation}\label{eq:NSlin}
\partial_t{\bf u}=-\frac{1}{\rho}\boldsymbol{\nabla}p-2\boldsymbol{\Omega}\times{\bf u} +\nu \nabla^2 {\bf
u},
\end{equation}
where ${\bf u}=(u_x,u_y,u_z)$ is the velocity, $p$ the pressure,
$\rho$ the fluid density, $\nu$ the kinematic viscosity, and
$\boldsymbol{\Omega} = \Omega {\bf e}_z$.

The perturbation induced by a local source can be formally
described by a localized field of rate of expansion $q({\bf x},t)$
acting via the continuity equation
\begin{equation}\label{eq:cont}
\boldsymbol{\nabla} \cdot {\bf u} = q.
\end{equation}
In the case of an object under translation at velocity ${\bf
U}=U{\bf e}_x$, one can write $q({\bf x},t)=q_0({\bf x}-U t {\bf
e}_x)$. Such representation has been introduced by
Miles~\cite{Miles1971} and Janowitz~\cite{Janowitz1984}, among
others, in the context of internal gravity waves;
Voisin~\cite{Voisin2007} discussed in this context the appropriate
expression for $q_0$ in the weak and strong stratification limits
in the case of a sphere.

Using combinations of derivatives of Eqs.~(\ref{eq:NSlin}) and
(\ref{eq:cont}), one can derive the equations of propagation of
inertial waves forced by $q$,
\begin{eqnarray}\label{eq:propagation1}
\left[\left(\frac{\partial}{\partial t} -\nu \nabla^2\right)^2
\nabla^2+(2\Omega)^2\frac{\partial^2}{\partial z^2}\right] u_x &=&
\left[\left(\frac{\partial}{\partial t} -\nu
\nabla^2\right)\frac{\partial}{\partial
x}+2\Omega\frac{\partial}{\partial
y}\right]\left(\frac{\partial}{\partial t} -\nu
\nabla^2\right)q,\\\label{eq:propagation2}
\left[\left(\frac{\partial}{\partial t} -\nu \nabla^2\right)^2
\nabla^2+(2\Omega)^2\frac{\partial^2}{\partial z^2}\right] u_y &=&
\left[\left(\frac{\partial}{\partial t} -\nu
\nabla^2\right)\frac{\partial}{\partial
y}-2\Omega\frac{\partial}{\partial
x}\right]\left(\frac{\partial}{\partial t} -\nu
\nabla^2\right)q,\\\label{eq:propagation3}
\left[\left(\frac{\partial}{\partial t} -\nu \nabla^2\right)^2
\nabla^2+(2\Omega)^2\frac{\partial^2}{\partial z^2}\right] u_z &=&
\left[\left(\frac{\partial}{\partial t} -\nu
\nabla^2\right)^2+(2\Omega)^2\right]\frac{\partial}{\partial z}q.
\end{eqnarray}

Considering a plane wave of wavevector ${\bf k}$ and angular
frequency $\sigma$ and equating the left-hand-side of
(\ref{eq:propagation1}) to zero leads to the viscous dispersion
relation of inertial waves
\begin{eqnarray}
\sigma &=& \sigma_r+i\sigma_i,\\
{\rm with}\quad \sigma_r &=& s 2\Omega \frac{k_z}{k}>0,\label{eq:invdisp}\\
{\rm and}\quad \sigma_i&=&\nu k^2,
\end{eqnarray}
where $\sigma_r$ and $\sigma_i$ are the real and imaginary parts
of $\sigma$ respectively, $k_z={\bf k}\cdot{\bf e}_z$, $k=|{\bf
k}|$ and $s=\sign(k_z)$. Fluid particles in such propagative
inertial wave describe circular translations along $-s{\bf k}$ in
planes normal to ${\bf k}$~\cite{GreenspanBook}. Because the wave
is transverse, vorticity ${\boldsymbol \omega}$, related to the
shear between planes of different phase, is parallel to the
velocity ${\bf u}$: the wave has helicity ${\boldsymbol
\omega}\cdot{\bf u}$, of sign given by $-s$. We note that such
inertial wave solution, derived here for small perturbations, is
also an exact solution of the full non-linear Navier--Stokes
equation. As a consequence, for finite ${\rm Ro}$ and outside of
viscous boundary layers, non-linearities can affect inertial waves
only via triadic interactions~\cite{Waleffe1993,Bordes2012}.

\subsection{Inviscid wake of a line object}\label{sec:line}

Before searching for solutions to
(\ref{eq:propagation1})--(\ref{eq:propagation3}), we first
determine the lines of constant phase of the linear inviscid wake
produced by the translation of a line object. The equations for
these lines have first been derived by Lighthill (see for
example~\cite{Lighthill1967}) and by Peat and
Stevenson~\cite{Peat1976} for the case of an arbitrary motion in a
fluid with both stratification and rotation. The method we use
here is identical to the one initially suggested by Lord
Kelvin~\cite{Kelvin1887} and Havelock~\cite{Havelock1908} to
describe the steady wake of surface gravity waves behind a ship
(see also~\cite{Lighthill1978,Darrigol2005}).

\begin{figure}
\centerline{\hspace{0.5cm}\includegraphics[width=9cm]{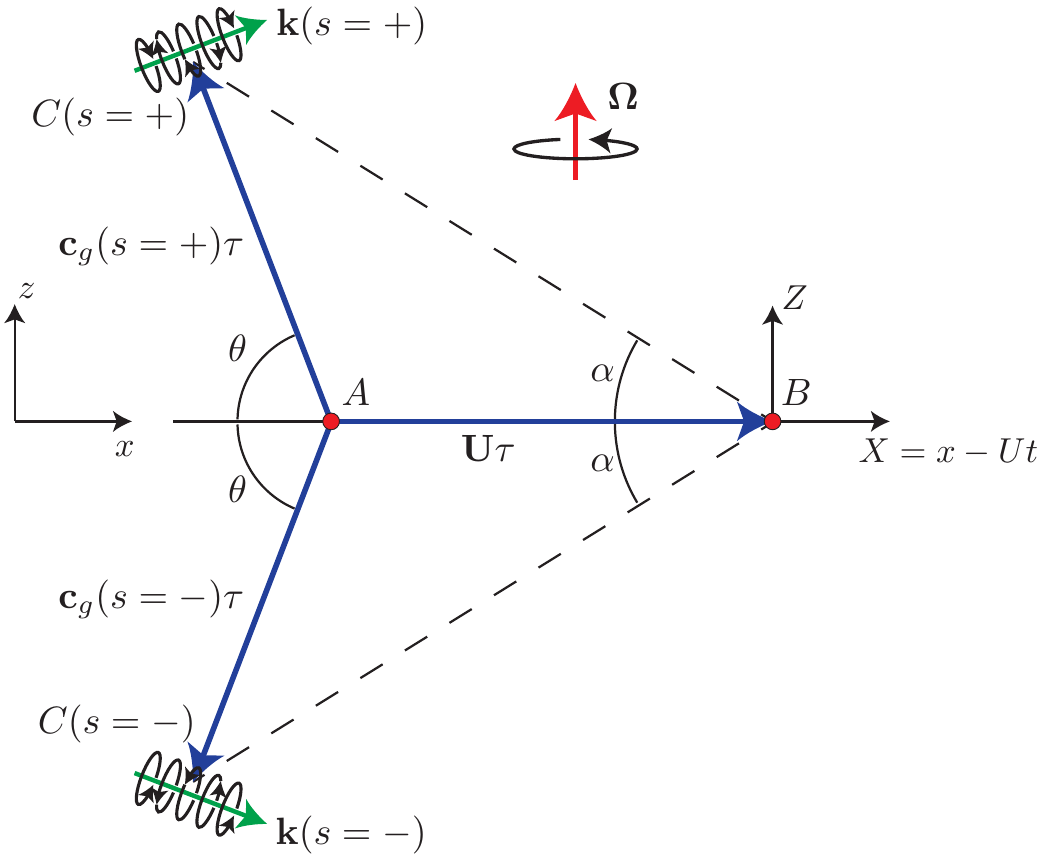}}
    \caption{Propagation of the two symmetric wavepackets, carrying
    wavenumber $k$, emitted at time $t-\tau$ by a horizontal line
    object (invariant along ${\bf e}_y$) in translation at constant
    velocity ${\bf U}=U{\bf e}_x$ in a fluid rotating at a rate
    $\boldsymbol{\Omega}=\Omega {\bf e}_z$. At time $t$, the line
    object is at point $B$, the upper wavepacket at point $C(s=+)$ and
    the lower wavepacket at $C(s=-)$. At time $t-\tau$, the line
    object was at point $A$. At locations $C(s=\pm)$, fluid particles
    describe a circular translation in planes tilted at an angle
    $\theta(k)$, with orientation given by the vector $-s{\bf k}(s)$.
    This circular translation motion propagates along wavevector ${\bf
    k}$ at the phase velocity $c_\varphi=\sigma/k=U\sin\theta$.
    ($X=x-Ut,Z=z$) is the system of coordinates attached to the moving
    object.}\label{fig:wake}
\end{figure}

In the linearized problem, a steady forcing produces a steady wake
in the frame moving with the disturbance. The wake can therefore
be described as a superposition of inertial waves that correspond
to steady perturbations in this moving frame. For a plane wave of
wavevector ${\bf k}=k(\sin\theta\, {\bf e}_x+s \cos\theta\,{\bf
e}_z)$ [see Fig.~\ref{fig:wake}], this stationarity condition
implies that the angular frequency $\sigma$ satisfies
\begin{eqnarray}\label{eq:stationnarity0}
\sigma={\bf k}\cdot{\bf U}=k U \sin\theta,
\end{eqnarray}
where $\theta\in[0;\pi/2]$. From the inviscid dispersion
relation~(\ref{eq:invdisp}), the stationarity condition becomes
\begin{eqnarray}\label{eq:stationnarity}
\tan\theta= \frac{2\Omega}{k U}=\frac{1}{{\rm Ro}_k},
\end{eqnarray}
where ${\rm Ro}_k$ is the Rossby number associated to wavenumber
$k$. This stationary condition associates a single angle $\theta$
to each wavenumber.

\begin{figure}
\centering \includegraphics[width=12cm]{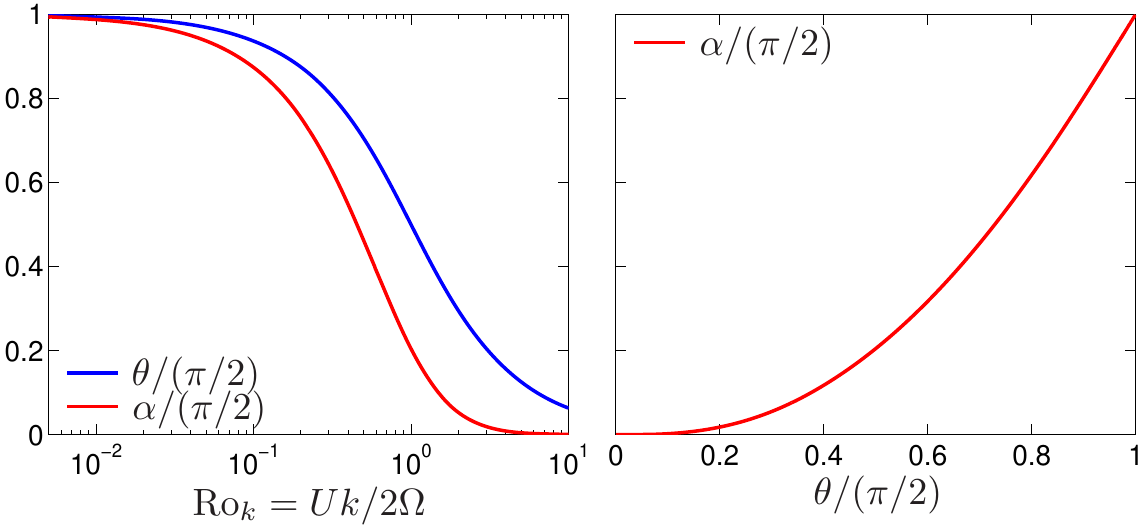}
    \caption{(a) Energy propagation angle in the frame of the fluid at
    rest $\theta$ (Eq.~\ref{eq:stationnarity}, blue line) and in the
    frame of the moving object $\alpha$ (Eq.~\ref{eq:alpha}, red line)
    as functions of the Rossby number ${\rm Ro}_k=Uk/2\Omega$ based on
    the wavenumber. (b)~$\alpha$ as a function of
    $\theta$.}\label{fig:angle}
\end{figure}

Assuming that the line object radiates inertial waves of any
wavenumber $k$, the position $C(k,\tau,s)$ of the wavepacket,
carrying wavenumber $k$, helicity sign $-s$ and produced at time
$t-\tau$, relative to the position $B$ of the line object at
current time $t$, is ${\bf BC}(k,\tau,s)= ({\bf c}_g - {\bf
U})\tau$, where
\begin{eqnarray}\label{eq:cg}
{\bf c}_g = \frac{2\Omega}{k}\sin\theta (-\cos\theta\,{\bf
e}_x+s\sin\theta\,{\bf e}_z)
\end{eqnarray}
is the group velocity associated to the wavevector ${\bf k}$ (see
Fig.~\ref{fig:wake})~\cite{GreenspanBook}. Using the stationarity
condition, the coordinates of $C(k,\tau,s)$ relative to the line
object at time $t$ can be written as
\begin{eqnarray}
    X(k,\tau,s)&=& -U \tau (1+\sin^2\theta),\label{eq:localx}\\
    Z(k,\tau,s)&=& s U \tau \frac{\sin^3\theta}{\cos\theta}.\label{eq:localy}
\end{eqnarray}
We can define the radiation angle $\alpha(k)$ as the angle along
which energy for a given wavenumber $k$ is supplied by the
disturbance in the moving frame. Writing
$\alpha=\tan^{-1}(|Z(k,\tau)/X(k,\tau)|)$ (see
Fig.~\ref{fig:wake}), this angle satisfies
\begin{eqnarray}\label{eq:alpha}
    \alpha=\tan^{-1}\left(\frac{\sin^3\theta}{\cos\theta(1+\sin^2\theta)}\right)=\tan^{-1}\left(\frac{1}{2{\rm Ro}_k+{\rm Ro}_k^3}\right).
\end{eqnarray}
The line at angle $\alpha(k)$ therefore corresponds to a line of
constant wavevector ${\bf k}$ (constant $\theta$ and $k$). The
radiation angle $\alpha(k)$ decreases monotonically with the
wavenumber $k$ from $90^\circ$ to $0^\circ$ (see
Fig.~\ref{fig:angle}a): smaller wavelengths are found closer to
the translation axis $z=0$. This contrasts with the case of
surface gravity waves, for which $\alpha(k)$ is maximum at a
finite wavenumber, which defines the famous Kelvin angle of ship
wakes ~\cite{Lighthill1978,Rabaud2013,Darmon2014}.

\begin{figure}
    \centerline{\includegraphics[width=9cm]{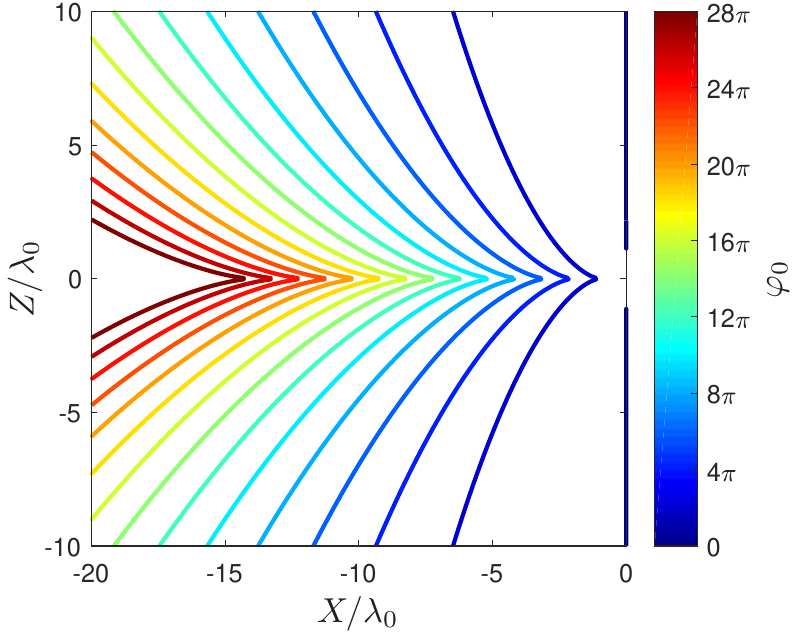}}
    \caption{Lines of constant phase $\varphi_0=2\pi n$ with
    $n\in[0:14]$ (Eqs.~\ref{eq:wake1}--\ref{eq:wake2}) in the wake of inertial
    waves of a line source (of axis ${\bf e}_y$) translated at
    velocity $U$ along ${\bf e}_x$ in an inviscid fluid under rotation
    at a rate $\Omega$ about ${\bf e}_z$. Coordinates are
    normalized by $\lambda_0=U\pi/\Omega$, the wake wavelength along
    the translation axis.} \label{fig:wakefield}
\end{figure}

The phase of the wave at point $C(k,\tau,s)$ can be computed as
\begin{eqnarray}
\varphi(k,\tau,s)= \varphi_e + \sigma \tau - {\bf k} \cdot {\bf
AC},
\end{eqnarray}
where $\varphi_e$ is the phase of the wave when it is emitted and
${\bf AC}={\bf c}_g \tau$ is the distance travelled by the
wavepacket between $t-\tau$ and $t$. The anisotropic dispersion
relation of inertial waves imposes that ${\bf k} \cdot {\bf
c}_g=0$, so that the equation for a line of constant phase
$\varphi=\varphi_0+\varphi_e$ satisfies
\begin{eqnarray}\label{eq:isophase1}
\varphi_0 = 2\Omega \tau \cos\theta.
\end{eqnarray}
Injecting (\ref{eq:isophase1}) in
(\ref{eq:localx})--(\ref{eq:localy}) finally provides a parametric
representation of the lines of constant phase,
\begin{eqnarray}
X&=&-\lambda_0 \frac{\varphi_0}{2\pi}\frac{1+\sin^2\theta}{\cos\theta},\label{eq:wake1}\\
Z&=&s\lambda_0
\frac{\varphi_0}{2\pi}\frac{\sin^3\theta}{\cos^2\theta}.\label{eq:wake2}
\end{eqnarray}
with $\lambda_0=\pi U/\Omega$ the wavelength of the wake along the
translation axis. In Fig.~\ref{fig:wakefield}(a), we plot the
lines of constant phase $\varphi_0(X,Z)=2\pi n$ for $n\in[0:14]$.
This phase pattern is identical to that predicted by
Lighthill~\cite{Lighthill1967} and Mowbray and
Rarity~\cite{Mowbray1967} for internal waves produced by the
vertical translation of a line or point source in a
density-stratified fluid (see also
Refs.~\cite{Stevenson1983,Gartner1986,Torres2000,Okino2017}). Note
that $\varphi_0$ is the phase relative to the phase at emission
$\varphi_e$ which remains unspecified here.

\subsection{Viscous wake of a finite size object}\label{sec:theory}

We now describe the wake of inertial waves of a translating object
of finite size, invariant along $y$, including viscosity effects.
For this, we integrate
(\ref{eq:propagation1})--(\ref{eq:propagation3}) accounting for
the geometry of the object by its equivalent field of rate of
expansion $q({\bf x},t)=q_0({\bf x}-Ut{\bf e}_x)$. The following
derivation leads to results similar to those in
Refs.~\cite{Stewartson1979,Johnson1982,Cheng1982,Heikes1982}, but
differs in the combination of approximations used: infinite depth,
free-slip boundary, weak streamwise perturbation, no
quasi-geostrophy. Their outcome will be compared to experimental
wakes in Sec.~\ref{sec:expresults}.

We first introduce the 3D spatio-temporal Fourier transform of the
field of rate of expansion $q({\bf x},t)$ invariant along $y$
\begin{eqnarray}\label{eq:fft}
\hat{q}({\bf k},\sigma)&=&\int q({\bf x},t)e^{i(\sigma t - {\bf
k}\cdot {\bf x})}\,dt\,dx\,dz,\\\label{eq:fftinv} q({\bf
x},t)&=&\frac{1}{(2\pi)^3}\int \hat{q}({\bf k},\sigma)e^{-i(\sigma
t - {\bf k}\cdot {\bf x})}\,d\sigma\,dk_x\,dk_z,
\end{eqnarray}
where ${\bf k}=(k_x,0,k_z)$. From
Eq.~(\ref{eq:propagation1})--(\ref{eq:propagation3}) and using the
relation $\hat{q}({\bf k},\sigma)=2\pi \widehat{q_0}({\bf
k})\delta(\sigma-{\bf k} \cdot {\bf U})=2\pi \widehat{q_0}({\bf
k})\delta(\sigma-k_x U)$, which accounts for the stationarity of
the forcing in the frame of the translating object, the velocity
field writes
\begin{eqnarray}
  \label{eq:vx2}
  u_x({\bf X}) & = & -\frac{i}{(2\pi)^2}\int
    \frac{(Uk_x+i\nu k^2)^2k_x}
         {(Uk_x+i\nu k^2)^2k^2-(2\Omega)^2k_z^2}
    \widehat{q_0}({\bf k})
    e^{i {\bf k}\cdot {\bf X}}
    \,dk_x\,dk_z,
    \\
  \label{eq:vy2}
  u_y({\bf X}) & = & -\frac{1}{(2\pi)^2}\int
    \frac{2\Omega (Uk_x+i\nu k^2)k_x}
         {(Uk_x+i\nu k^2)^2k^2-(2\Omega)^2k_z^2}
    \widehat{q_0}({\bf k})
    e^{i {\bf k}\cdot {\bf X}}
    \,dk_x\,dk_z,
    \\
  \label{eq:vz2}
  u_z({\bf X}) & = & -\frac{i}{(2\pi)^2}\int
    \frac{[(Uk_x+i\nu k^2)^2 -(2\Omega)^2]k_z}
         {(Uk_x+i\nu k^2)^2k^2-(2\Omega)^2k_z^2}
    \widehat{q_0}({\bf k})
    e^{i {\bf k}\cdot {\bf X}}
    \,dk_x\,dk_z,
\end{eqnarray}
with ${\bf X}={\bf x}-{\bf U}t$ the position relative to the
object.

Let us introduce the Rossby number ${\rm Ro}_{k_x} = Uk_x/2\Omega$
and the Reynolds number ${\rm Re}_{k_x} = U/\nu k_x$ associated
with the horizontal wavenumber $k_x$. Application of the residue
theorem to the integral over $k_z$ allows its evaluation in the
small viscosity limit ${\rm Re}_{k_x} \gg 1$. The integrand has
three poles in the half-plane $s\Im(k_z)
> 0$, with $s = \sign(z)$, picked by Jordan's lemma. The first two
poles,
\begin{equation}\label{eq:pole_Ek}
  k_z
  = s\left(\frac{Uk_x\pm2\Omega}{\nu}\right)^{1/2}e^{i\pi/4}
  = s\sqrt{\frac{2\Omega}{\nu}}({\rm Ro}_{k_x}\pm1)^{1/2}e^{i\pi/4},
\end{equation}
where
\begin{subequations}\label{eq-detrootlayer}
\begin{alignat}{2}
  (Uk_x\pm 2\Omega)^{1/2}
  & =
    \sqrt{|Uk_x\pm 2\Omega|}
    & \qquad & \text{when $Uk_x > \mp 2\Omega$},
    \\
  & =
    i\sqrt{|2\Omega\pm Uk_x|}
    & & \text{when $Uk_x < \mp 2\Omega$},
\end{alignat}
\end{subequations}
correspond to the boundary layer along the object: we can note
that the modulus of these two poles (\ref{eq:pole_Ek}) tends
toward the inverse of the thickness $\sqrt{\nu/2\Omega}$ of an
Ekman layer when the Rossby number ${\rm Ro}_{k_x}$ vanishes. The
third pole
\begin{equation}
  k_z = k_{z,r}(k_x)+ik_{z,i}(k_x),
  \label{eq-wavpolfull}
\end{equation}
of inviscid real part
\begin{equation}
  \label{eq:kz}
  k_{z,r}
   = s
     \frac{U k_x |k_x|}
          {[(2\Omega)^2-(Uk_x)^2]^{1/2}}
   = s
     \frac{|k_x|}
          {(1/{\rm Ro}_{k_x}^2-1)^{1/2}},
\end{equation}
and viscous imaginary part
\begin{equation}
  \label{eq-wavpolvisc}
  k_{z,i}
  = s
    \frac{(2\Omega)^4\nu k_x^2|k_x|}
         {[(2\Omega)^2-(Uk_x)^2]^{5/2}}
  = \frac{s}{{\rm Re}_{k_x}}
    \frac{|k_x|}
         {{\rm Ro}_{k_x}^4(1/{\rm Ro}_{k_x}^2-1)^{5/2}},
\end{equation}
where
\begin{subequations}\label{eq-detrootwaves}
\begin{alignat}{2}
  [(2\Omega)^2-(Uk_x)^2]^{1/2}
  & =
    \sqrt{|(2\Omega)^2-(Uk_x)^2|}
    & \qquad & \text{when $U|k_x| < 2\Omega$},
    \\
  & =
    -\mathrm{i}\sqrt{|(Uk_x)^2-(2\Omega)^2|}\sign(k_x)
    & & \text{when $U|k_x| > 2\Omega$},
\end{alignat}
\end{subequations}
corresponds to waves slowly dissipated by viscosity as they
propagate away from the object. We can note that the spatial decay
factor of the wake component associated to $k_x$
\begin{equation}\label{eq:decayfact}
k_{z,i}Z = s \frac{\nu}{2\Omega} \frac{k_r^5}{k_x^2}Z,
\end{equation}
where $k_r^2=k_x^2+k_{z,r}^2$ can be written as $\nu k_r^2 \tau$,
where $\tau$ is the time for the wavepacket carrying wavevector
$(k_x,k_{z,r})$ to travel from the object to ${\bf X}$ at the
group velocity. This viscous decay could already have been derived
in Sec.~\ref{sec:line} from the temporal decay due to viscosity of
the wave amplitude in the wavepacket carrying wavenumber $k_r$.
The determination of the complex square roots
(\ref{eq-detrootlayer}) and (\ref{eq-detrootwaves}) may be seen as
the insertion, in the complex $k_x$-plane, of branch cuts
extending from the singularities $k_x = \pm 2\Omega/U$ vertically
downwards.

We can note that Eq.~(\ref{eq:kz}) is identical to the wave
stationarity condition~(\ref{eq:stationnarity}), with $\sign(z)$
playing here the same role as $\sign(k_z)$ there. From
Eq.~(\ref{eq:kz}), it is apparent that horizontal wavenumbers
$k_x$ larger than $k_0=2\Omega/U$ in absolute value will not
contribute significantly to the wake of inertial waves since, to
leading order in $1/{\rm Re}_{k_x}$, the associated vertical
wavenumber $k_z$ is imaginary and leads to vertically evanescent
waves. It is worth noting that, contrary to Stewartson and
Cheng~\cite{Stewartson1979}, Johnson~\cite{Johnson1982} and Cheng
and Johnson~\cite{Cheng1982}, but as in Heikes and
Maxworthy~\cite{Heikes1982}, no quasi-geostrophic approximation of
weak vertical derivatives $k_z \ll k_x$ (i.e. ${\rm Ro}_{k_x}=U
k_x/2\Omega \ll 1$) is made here. The only approximations up to
now are the linearization of the Navier--Stokes equation by
assuming small velocity perturbations and small viscosity.

In the experiments reported in section~\ref{sec:expresults}, we
consider the translation of a cylinder of diameter $d$ associated
to Reynolds numbers ${\rm Re}=Ud/\nu$ typically ranging from $10$
to $1000$ and Rossby numbers ${\rm Ro}=U/2\Omega d$ ranging from
$10^{-2}$ to $1$. The Ekman boundary layer on the cylinder, of
typical thickness $\sqrt{\nu/2\Omega}=d\sqrt{{\rm Ro}/{\rm Re}}$,
is therefore expected to remain small compared to the cylinder for
most of our experiments, and the contribution of the corresponding
poles in the integration of (\ref{eq:vx2})--(\ref{eq:vz2}) will be
neglected in the following. Retaining only the pole
(\ref{eq-wavpolfull})--(\ref{eq-wavpolvisc}), the velocity field
follows as
\begin{eqnarray}
  \label{eq:vx3}
  u_x & = &
    -\frac{1}{4\pi}\int
    \frac{U|k_x|}
         {[(2\Omega)^2-(Uk_x)^2]^{1/2}}
    \widehat{q_1}(k_x)
    e^{-k_{z,i}(k_x)Z}
    e^{i[k_x X+k_{z,r}(k_x)Z]}
    \,dk_x,
    \\
  \label{eq:vy3}
  u_y & = &
    \frac{i}{4\pi}\int
    \frac{2\Omega \sign(k_x)}
         {[(2\Omega)^2-(Uk_x)^2]^{1/2}}
    \widehat{q_1}(k_x)
    e^{-k_{z,i}(k_x)Z}
    e^{i[k_x X+k_{z,r}(k_x)Z]}
    \,dk_x,
    \\
  \label{eq:vz3}
  u_z & = &
    \frac{\sign(Z)}{4\pi}\int
    \widehat{q_1}(k_x)
    e^{-k_{z,i}(k_x)Z}
    e^{i[k_x X+k_{z,r}(k_x)Z]}
    \,dk_x,
\end{eqnarray}
where $X=x-Ut$, $Z=z$ and
$\widehat{q_1}(k_x)=\widehat{q_0}(k_x,k_{z,r}(k_x))$.

\subsubsection{Far-field limit}
In the far field limit $k_0|{\bf X}| \gg 1$, the integrals over
$k_x$~(\ref{eq:vx3})--(\ref{eq:vz3}) may be evaluated
asymptotically. In the inviscid case, the phase of the integrand
is real in the range $|k_x| < k_0$ of propagating waves and the
stationary phase method may be applied (see
Appendix~\ref{appendix1}). The waves are only found downstream, in
the half-plane $X < 0$. There, two opposite stationary points $k_x
= \pm k_0\cos\theta$ are obtained for each radiation angle
$\alpha=\tan^{-1}(|Z/X|)$, the angle $\theta$ satisfying the cubic
equation
\begin{equation}
  \cot^3\theta+2\cot\theta-\cot\alpha = 0,
  \label{eq-cubicequa}
\end{equation}
consistent with (\ref{eq:alpha}), of real root
\begin{equation}
  \cot\theta =
  \left(
    \sqrt{\frac{\cot^2\alpha}{4}+\frac{8}{27}}
    +\frac{\cot\alpha}{2}
  \right)^{1/3}-
  \left(
    \sqrt{\frac{\cot^2\alpha}{4}+\frac{8}{27}}
    -\frac{\cot\alpha}{2}
  \right)^{1/3}.
  \label{eq-theta}
\end{equation}
The associated wavevectors are $\pm{\bf k}_s$, with
\begin{equation}
  {\bf k}_s =
  k_0\cot\theta
  (\sin\theta{\bf e}_x+s\cos\theta{\bf e}_z),
\end{equation}
consistent with (\ref{eq:stationnarity}), and we also recover the
same phase field as the one described in Sec.~\ref{sec:line}.

Viscosity adds a small imaginary part to the phase of the
integrand in (\ref{eq:vx3})--(\ref{eq:vz3}). To leading order in
$1/{\rm Re}$ (see Appendix~\ref{appendix1}), the result is a slow
exponential decay as
$\exp(-k_{z,i}Z)=\exp[-k_0|Z|\cos^3\theta/({\rm Re}\,{\rm
Ro}\sin^5\theta)]$. The far-field velocity follows as
\begin{align}
  u_x & = -H(-X)
    \frac{\sin^{3/2}\theta\cos\theta}{\sqrt{2+\cos^2\theta}}
    \exp
    \left(
      -\frac{k_0|Z|}{{\rm Re}{\rm Ro}}
      \frac{\cos^3\theta}{\sin^5\theta}
    \right)
    \frac{\Re\left[\widehat{q_0}({\bf k}_s)e^{-i\varphi_s}\right]}
         {\sqrt{\lambda_0|Z|}},
    \label{eq-farvx}
    \\
  u_y & = -H(-X)
    \frac{\sin^{3/2}\theta}{\sqrt{2+\cos^2\theta}}
    \exp
    \left(
      -\frac{k_0|Z|}{{\rm Re}{\rm Ro}}
      \frac{\cos^3\theta}{\sin^5\theta}
    \right)
    \frac{\Im\left[\widehat{q_0}({\bf k}_s)e^{-i\varphi_s}\right]}
         {\sqrt{\lambda_0|Z|}},
    \label{eq-farvy}
    \\
  u_z & = H(-X)
    \frac{\sin^{5/2}\theta\sign(Z)}{\sqrt{2+\cos^2\theta}}
    \exp
    \left(
      -\frac{k_0|Z|}{{\rm Re}{\rm Ro}}
      \frac{\cos^3\theta}{\sin^5\theta}
    \right)
    \frac{\Re\left[\widehat{q_0}({\bf k}_s)e^{-i\varphi_s}\right]}
         {\sqrt{\lambda_0|Z|}},
    \label{eq-farvz}
\end{align}
where $H$ denotes the Heaviside step function, $\Re$ and $\Im$ the
real and imaginary parts respectively, and
\begin{equation}
  \varphi_s =
  \underbrace{k_0|Z|\frac{\cos^2\theta}{\sin^3\theta}}_{=\varphi_0}
  -\underbrace{\frac{\pi}{4}}_{=\varphi_e}.
  \label{eq-farphaseinv}
\end{equation}
In this expression, the first term
$k_0|Z|\cos^2\theta/\sin^3\theta$ identifies to the phase field
$\varphi_0$ derived in section~\ref{sec:line}, while the second
term allows us to identify the value of the phase at emission,
$\varphi_e=\pi/4$, which was left unknown in
Section~\ref{sec:line}. The lines of constant phase
$\varphi_s=\text{cst}$ have the parametric equation
\begin{align}
  k_0|X| & = \frac{1+\sin^2\theta}{\cos\theta}
    \left(
      \varphi_s+\frac{\pi}{4}
    \right),
    \label{eq:viswake1}
    \\
  k_0|Z| & = \frac{\sin^3\theta}{\cos^2\theta}
    \left(
      \varphi_s+\frac{\pi}{4}
    \right),
    \label{eq:viswake2}
\end{align}
consistent with (\ref{eq:wake1})--(\ref{eq:wake2}).
Johnson~\cite{Johnson1982} pointed out that, as $\rm Re$ decreases
to order unity, viscosity also induces a deformation of these
lines. This effect is discussed in Appendix~\ref{appendix2}.

\subsubsection{Boundary condition, model for the object
spectrum}\label{sec:modelcylinder}

We finally introduce a model for the spectrum $\widehat{q_1}(k_x)$
of the equivalent source of inertial waves induced by the motion
of the object. Since we neglect here the Ekman boundary layer on
the object surface, we use an inviscid non-penetration boundary
condition as in Refs.~\cite{Stewartson1979,Cheng1982,Heikes1982},
\begin{eqnarray}
({\bf u}-{\bf U})\cdot {\bf n}=0, \label{eq-genbc}
\end{eqnarray}
where ${\bf n}$ is the vector normal to the object surface and
${\bf u}-{\bf U}$ the fluid velocity in the reference frame of the
object. We consider in the following a symmetric object, of
boundary described by $Z=\pm f(X)$, yielding a boundary condition
\begin{equation}
  \label{eq:bc}
  u_z = \pm (u_x-U)
  \frac{d f}{d X}.
\end{equation}

We are here mainly interested in wakes at small Rossby number, for
which finite size effects are expected (${\rm Ro}=U/2\Omega d$
compares the size $d$ of the object to the characteristic
wavelength $\lambda_0=\pi U/\Omega$ of the inviscid wake for a
line object). In this regime, the horizontal wavenumbers of the
order of $k_x \sim 2\pi/d$, which are associated to small ${\rm
Ro}_{k_x}=Uk_x/2\Omega$, are expected to dominate the wake. From
Eq.~(\ref{eq:kz}), the leading spectral components verify
$|k_{z,r} Z| \simeq {\rm Ro}_{k_x}|k_x Z|\ll 1$ for $|Z|\leq d$.
Accordingly, the phase of the dominant components in the integrand
of (\ref{eq:vx3})--(\ref{eq:vz3}) shows weak variations over the
vertical extent $|Z|\leq d$ of the object, so it is justified to
apply the boundary condition (\ref{eq:bc}) at $Z=0_\pm$ instead of
$Z=\pm f(X)$.

In the following, we consider an object invariant along $y$, of
aspect ratio of order 1 in the vertical plane. Still assuming
small Rossby number ${\rm Ro}=U/2\Omega d$, the stationarity
condition~(\ref{eq:stationnarity}) (or equivalently
Eqs.~\ref{eq:kz} and \ref{eq:vx3}--\ref{eq:vz3}) suggests that the
ratio $u_x/u_z$ of the axial to vertical velocity perturbations
due to the leading spectral terms of the wake, which is given by
$1/\tan\theta={\rm Ro}_k\simeq {\rm Ro}_{k_x}$, is small as well.
We will therefore assume in~Eq.~(\ref{eq:bc}) that the horizontal
velocity $u_x$ is negligible with respect to $u_z$ close to the
cylinder, $u_z$ being itself assumed to be of order $U$ because of
the aspect ratio $1$ of the object: this constitutes the
assumption of weak streamwise perturbation at the core of the
present model.

In a non-rotating fluid, the approximation $u_x \ll U$ is relevant
only for slender bodies verifying $d f/d X \ll 1$. Here, it also
applies for a bluff body because of the small value of the Rossby
number, provided that the body is invariant along $y$, as first
noted by Johnson~\cite{Johnson1982} and Heikes and
Maxworthy~\cite{Heikes1982}. This can be understood by eliminating
the pressure term in the expressions of the 4D spatio-temporal
Fourier transform of the inviscid and linearized Navier--Stokes
equation, which leads to
\begin{eqnarray}
\widehat{u_x} &=&  i \frac{k_z}{k}\widehat{u_y} - i \frac{k_y}{k}\widehat{u_z},\label{eq:FFTNS2}\\
\widehat{u_y} &= & -i \frac{k_{z}}{k}\widehat{u_x} + i
\frac{k_x}{k}\widehat{u_z}.\label{eq:FFTNS3}
\end{eqnarray}
For small Rossby numbers ${\rm Ro}_{k_x}$, the wave stationarity
condition gives $k_z/k \simeq {\rm Ro}_{k_x}$ and $k_x/k \simeq
O(1)$. Equations~(\ref{eq:FFTNS2})--(\ref{eq:FFTNS3}) therefore
show that $u_x/u_z=O({\rm Ro}_{k_x})$ if $k_y = 0$, whereas $k_y
\sim k_x$ and $u_x/u_z=O(1)$ for a 3D object of aspect ratio of
order $1$.

Under these approximations, the boundary condition (\ref{eq:bc})
simply writes
\begin{eqnarray}
u_z =\mp U \frac{d f}{d X} \quad {\rm at}
  \quad Z=0_\pm.
\end{eqnarray}
This velocity discontinuity implies a rate of expansion
\begin{equation}
  q_0(X,Z)
  = [u_z(X,Z=0_+)-u_z(X,Z=0_-)]\delta(Z)
  = -2 U f'(X)\delta(Z)
  \label{eq:bcq}
\end{equation}
leading to
\begin{equation}
  q_1(X) = -2 U f'(X),
  \label{eq:bcq2}
\end{equation}
which is consistent with Eq.~(\ref{eq:vz3}) applied at $Z=0_\pm$.
In the following, we consider a cylinder of diameter $d=2R$ such
that $f(X)=H(R-|X|)\sqrt{R^2-X^2}$, where $H$ is the Heaviside
step function. The boundary condition writes
\begin{equation}
  u_z(X,Z=0_\pm) = \pm U\frac{X}{\sqrt{R^2-X^2}}H(R-|X|),
  \label{eq:bccylvel}
\end{equation}
yielding the representation
\begin{equation}
  \label{eq:bccylq}
  q_1(X) = 2U\frac{X}{\sqrt{R^2-X^2}} H(R-|X|),
  \qquad
  \widehat{q_1}(k_x) =
    - 2i\pi R U J_1(k_x R),
\end{equation}
where $J_1$ is the Bessel function of the first kind of order 1.

\begin{figure}
    \centerline{\includegraphics[width=14cm]{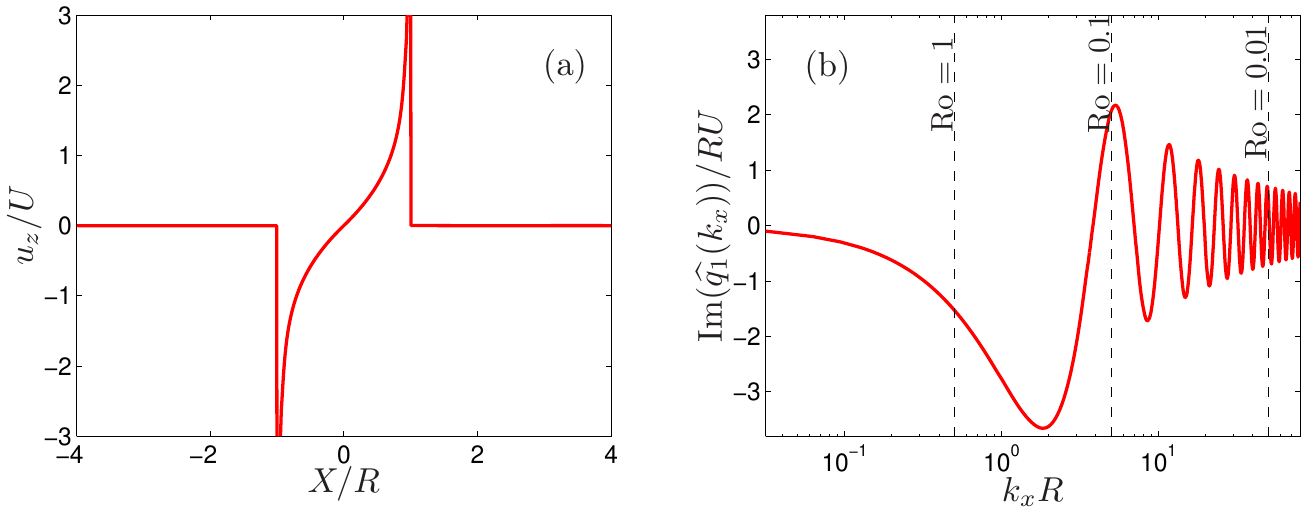}}
    \caption{(a) Vertical velocity profile $u_z(X)/U$ at $Z=0_+$ used
    as the boundary condition (Eq.~\ref{eq:bccylvel}) in the
    weak-streamwise-perturbation approximation. (b) Corresponding
    spectrum $\Im(\widehat{q_1}(k_x))/RU$ (Eq.~\ref{eq:bccylq}). We
    report with vertical dashed lines the critical horizontal
    wavevector component $k_0=1/{\rm Ro}\,d$ above which waves are
    vertically evanescent for three values (${\rm Ro}=0.01$, $0.1$ and
    $1$) of the Rossby number ${\rm Ro}=U/2\Omega
    d$.}\label{fig:theoprofile}
\end{figure}

Fig.~\ref{fig:theoprofile} shows the vertical velocity at the
boundary (\ref{eq:bccylvel}), and the corresponding spectrum
$\widehat{q_1}(k_x)$ (\ref{eq:bccylq}). In
Fig.~\ref{fig:theoprofile}(b), we show with vertical dashed lines
the critical horizontal wavevector component $k_0=1/({\rm Ro}\,d)$
above which waves are vertically evanescent for three values of
the Rossby number (${\rm Ro}=0.01$, $0.1$ and $1$): only the
portion of the spectrum $\widehat{q_1}(k_x)$ at wavenumbers $k_x$
smaller than $k_0$ contributes to the wake.

\begin{figure}
    \centerline{\includegraphics[width=\textwidth]{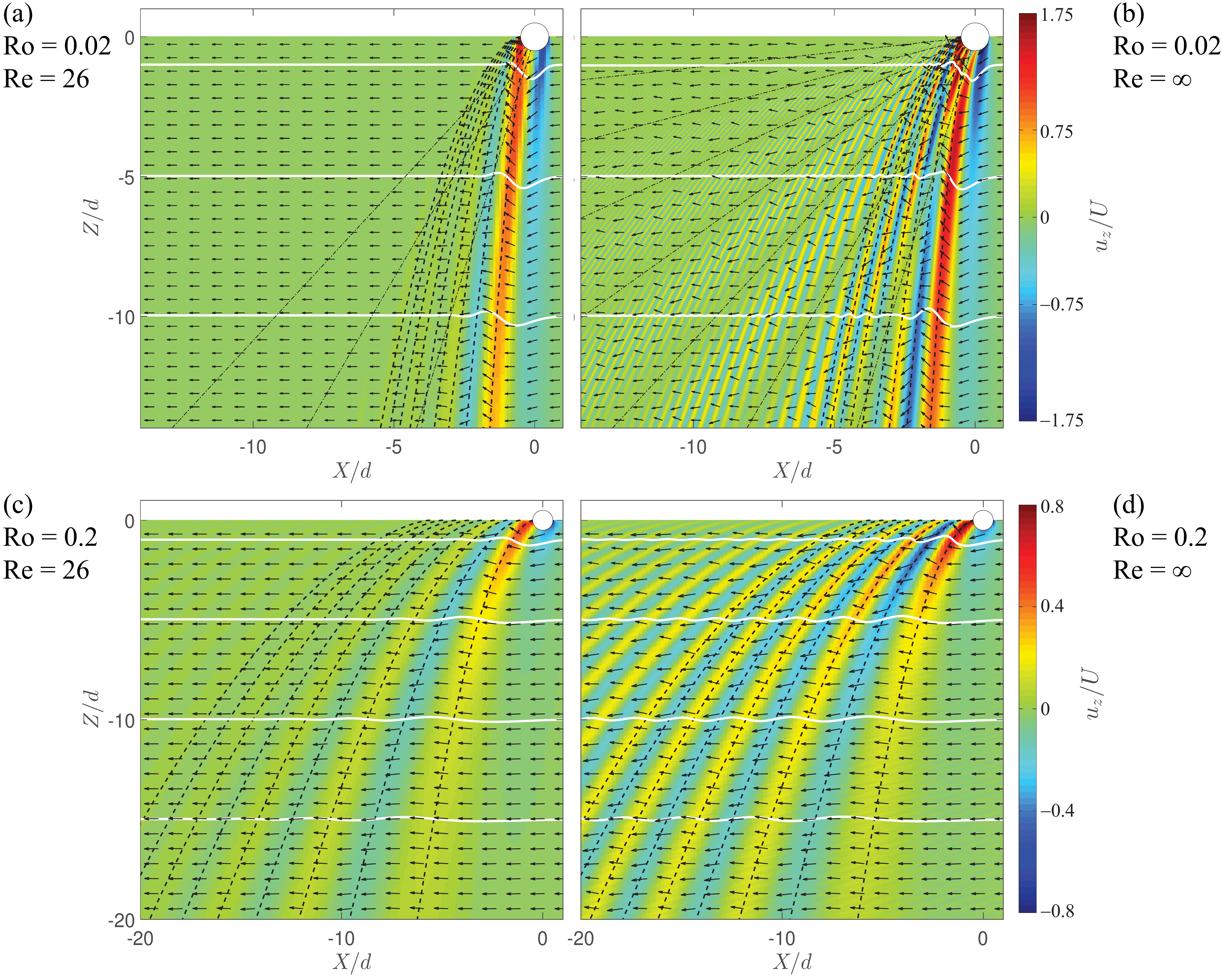}}
\caption{Wake structure computed from the
weak-streamwise-perturbation model (\ref{eq:vx3})--(\ref{eq:vz3})
for a cylinder of diameter $d$, invariant along the direction $y$.
Arrows show the in-plane velocity components $(u_x,u_z)$ in the
reference frame of the cylinder, and the colormap the vertical
velocity component $u_z$ normalized by $U$. White lines show the
projection of streamlines in the vertical plane. In (a) and (b),
the dashed-dotted lines show the radiation angles $\alpha_{\rm
zero}^n$, associated to the first roots $k_{x,{\rm zero}}^n R$ of
the spectrum, along which no energy is present. In (c) and (d),
these angles of zero amplitude do not exist since the roots of the
spectrum $\widehat{q_1}(k_x)$ correspond to evanescent waves. In
(a--d), we also show with dashed lines a few lines of constant
phase of the far-field wake (\ref{eq-farvx}-\ref{eq-farvz}). We
actually plot the parametric
curve~(\ref{eq:viswake1}-\ref{eq:viswake2}) for $\varphi_s+\pi/2 =
\pi$, $2\pi$, $3\pi$, $4\pi$, \ldots, $9\pi$ (from right to left)
corresponding to the local maxima and minima of $u_z$ in the
far-field theory (Eq.~\ref{eq-farvz}) for the
spectrum~(\ref{eq:bccylq}) (the spectrum~(\ref{eq:bccylq}) being
imaginary introduces the $+\pi/2$ phase shift).}
\label{fig:slender}
\end{figure}

The wake structure, computed from
Eqs.~(\ref{eq:vx3})--(\ref{eq:vz3}) using a FFT algorithm, is
shown in Fig.~\ref{fig:slender} for two Rossby numbers, ${\rm
Ro}=0.02$ and $0.20$, and two Reynolds numbers, ${\rm Re}=26$ and
$\infty$. In the low Rossby number case
[Fig.~\ref{fig:slender}(b)], we observe the concentration of
energy along a set of radiation angles $\alpha_{\rm extr}^n$
corresponding to the extrema of the Bessel function ($k_{x,{\rm
extr}}^n R\simeq 1.84$, $5.33$, $8.54$, \ldots), separated by
angles $\alpha_{\rm zero}^n$ along which no energy is present
(highlighted with dashed-dotted lines) corresponding to its roots
($k_{x,{\rm zero}}^n R\simeq 3.83$, $7.02$, $10.17$, \ldots).
These oscillations are the interference pattern due to the finite
size of the object. According to Eqs.~(\ref{eq:alpha}) and
(\ref{eq:kz}), the specific angles $\alpha_{\rm extr}^n$ and
$\alpha_{\rm zero}^n$ decrease with the order $n$ and with the
cylinder Rossby number $\rm Ro$. In Fig.~\ref{fig:slender}(b), in
each angular sector between successive $\alpha_{\rm zero}^n$, we
observe a good agreement between the computed velocity field and
the lines of constant phase~(\ref{eq:viswake1}-\ref{eq:viswake2})
predicted for a line object: we however observe a sign change of
the velocity perturbation at each $\alpha_{\rm zero}^n$ along the
line of constant phase of the line object associated to the
corresponding sign change of $\widehat{q_1}(k_x)$. Looking now at
Fig.~\ref{fig:slender}(a) for the same Rossby number as in (b) but
for ${\rm Re}=26$, we see that, except for the beam corresponding
to the first and largest minimum of $\widehat{q_1}(k_x)$ (at
$k_{x,{\rm extr}}^1 R \simeq 1.84$), all the beams are rapidly
damped by viscosity, leading to a wake of typical angle
$\alpha_{\rm extr}^1$($\simeq 82^{\circ}$ here for ${\rm
Ro}=0.02$) and to a significant discrepancy with the lines of
constant phase of the line object.

For ${\rm Ro}={\rm Ro}_c\equiv 1/(2 k_{x,{\rm zero}}^1 R)\simeq
0.13$, the radiation angle $\alpha_{\rm zero}^1$ associated with
the first zero of the spectrum reaches $0^{\circ}$, while the zero
itself $k_{x,{\rm zero}}^1 R\simeq 3.83$ matches the critical
wavenumber $k_0 R=1/2{\rm Ro}$ above which inertial waves are
evanescent. As $\rm Ro$ further increases above ${\rm Ro}_c$, a
smaller portion of the spectrum $k_x<k_0$ ($<k_{x,{\rm zero}}^1$)
contributes to the wake, containing no zero. Considering for
example the wake at Rossby number ${\rm Ro}=0.20>{\rm Ro}_c$ in
Figs.~\ref{fig:slender}(c,d), we see that the two Reynolds numbers
considered here, ${\rm Re}=26$ and $\infty$, lead this time to
similar wake structures which match well the lines of constant
phase for a line object~(\ref{eq:viswake1}-\ref{eq:viswake2}). We
can conclude that, in this weak-streamwise-perturbation model,
${\rm Ro}_c=1/k_{x,{\rm zero}}^1 d \simeq 0.13$ stands as an
approximate threshold for the appearance of finite size effects in
the wake of a cylinder. We finally note that, in the finite-${\rm
Re}$ cases of Figs.~\ref{fig:slender}(a) and (c), the wake is
damped by viscosity more efficiently as the wake angle
$\alpha=\tan^{-1}(|Z/X|)$ becomes smaller. Recalling that the
decay factor for a propagating wave is $\exp(-\nu k^2 \tau)$, this
result is consistent with the prediction of a decreasing
wavenumber $k$ with $\alpha$ (Eq.~\ref{eq:alpha}) made in
Sec.~\ref{sec:line} for an infinitely small object. We also note
that the wake in Figs.~\ref{fig:slender}(a) and (c) is
progressively damped along a given direction $\alpha$, in
agreement with the fact that it corresponds to an increasing
propagation time $\tau$ at constant wavenumber $k$.

\begin{figure}
    \centerline{\includegraphics[width=\textwidth]{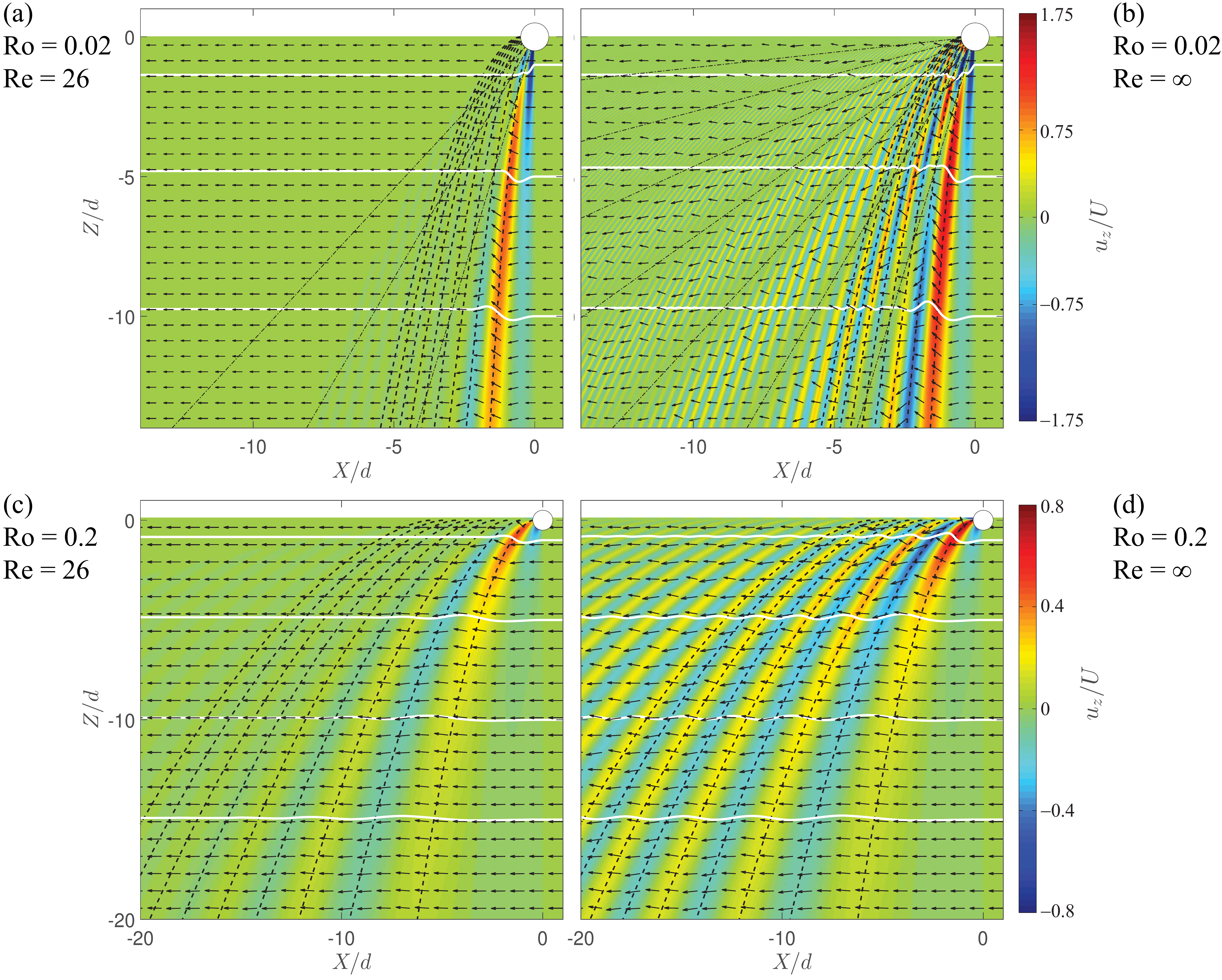}}
    \caption{Wake structure, for the same values of $\rm Ro$ and $\rm
    Re$ as in Fig.~\ref{fig:slender}, predicted in the far-field
    approximation (\ref{eq-farvx}-\ref{eq-farvz}) of the
    weak-streamwise-perturbation model (spectrum~\ref{eq:bccylq}) for
    a cylinder of diameter $d$, invariant along the direction $y$. The
    layout is the same as in Fig.~\ref{fig:slender}.}
    \label{fig:farfield}
\end{figure}

Finally, for the sake of comparison with the full
model~(\ref{eq:vx3}-\ref{eq:vz3}), we show in
Fig.~\ref{fig:farfield} the velocity field predicted in the
far-field approximation (\ref{eq-farvx}-\ref{eq-farvz}), still
using the weak-streamwise-perturbation spectrum of the
cylinder~(\ref{eq:bccylq}), for the same values of $\rm Ro$ and
$\rm Re$ as in Fig.~\ref{fig:slender}. One can see that for $\rm
Ro=0.2$ the far-field wake in Figs.~\ref{fig:farfield}(c-d) and
the one of the full model in Figs.~\ref{fig:slender}(c-d) are
almost identical for distances from the cylinder larger than
typically its diameter. On the other hand, comparing now
Figs.~\ref{fig:farfield}(a-b) and Figs.~\ref{fig:slender}(a-b) at
${\rm Ro}=0.02$ reveals significant differences up to distances
much larger than the cylinder diameter, with a much thinner ``main
beam'' ---the nearly vertical beam found around the wake angle
$\alpha_{\rm extr}^1 \simeq 82^{\circ}$. We note in particular
that the absence of flow perturbation upstream, that is for $X>0$,
in the far-field model constitutes a significant discrepancy with
the full model at low $\rm Ro$.

The absence of upstream flow perturbation in
Fig.~\ref{fig:farfield} is a manifestation of the nonuniformity of
the expansion~(\ref{eq-farvx}-\ref{eq-farvz}) at the wavefront $X
= 0$: there, diffraction takes place, requiring the switch to a
uniform expansion valid for all $X$ and involving Fresnel
functions. The mathematical origin of the nonuniformity and the
derivation of the uniform far-field expansion are briefly
discussed in Appendix~\ref{sec-uniform}. The uniform expansion
gives to the wave field some extension upstream, however still
smaller than with the full model. Another noticeable feature of
expansion~(\ref{eq-farvx}-\ref{eq-farvz}), visible in
Fig.~\ref{fig:farfield}, is an unphysical vertical shift of the
streamlines between their original position upstream and their
final position downstream. The uniform far-field
expansion~(\ref{eq-uxfaruni}-\ref{eq-uzfaruni}) presented in
Appendix~\ref{sec-uniform} reduces the vertical shift for most
streamlines (see Fig.~\ref{fig-unifarfield}), however not for
those close to the cylinder at low $\rm Ro$ (i.e. for ${\rm
Ro}=0.02$). The origin of this vertical shift of the streamlines
is related to a wrong estimation by the far-field expansion of the
velocity perturbation in the ``near field'' region close to the
cylinder (which increases in size as ${\rm Ro}$ decreases) as well
as in the region close to the axis $X=0$ for the non-uniform
expansion~(\ref{eq-farvx}-\ref{eq-farvz}).

Overall, the comparison of Figs.~\ref{fig:slender} and
\ref{fig:farfield} shows that the full
model~(\ref{eq:vx3}-\ref{eq:vz3}) should be preferred to fully
describe the cylinders' wake for Rossby numbers below typically
$0.1$.

\section{Experiments}\label{sec:expresults}

\subsection{Experimental setup}\label{sec:setup}

\begin{figure}
    \centerline{\includegraphics[width=12cm]{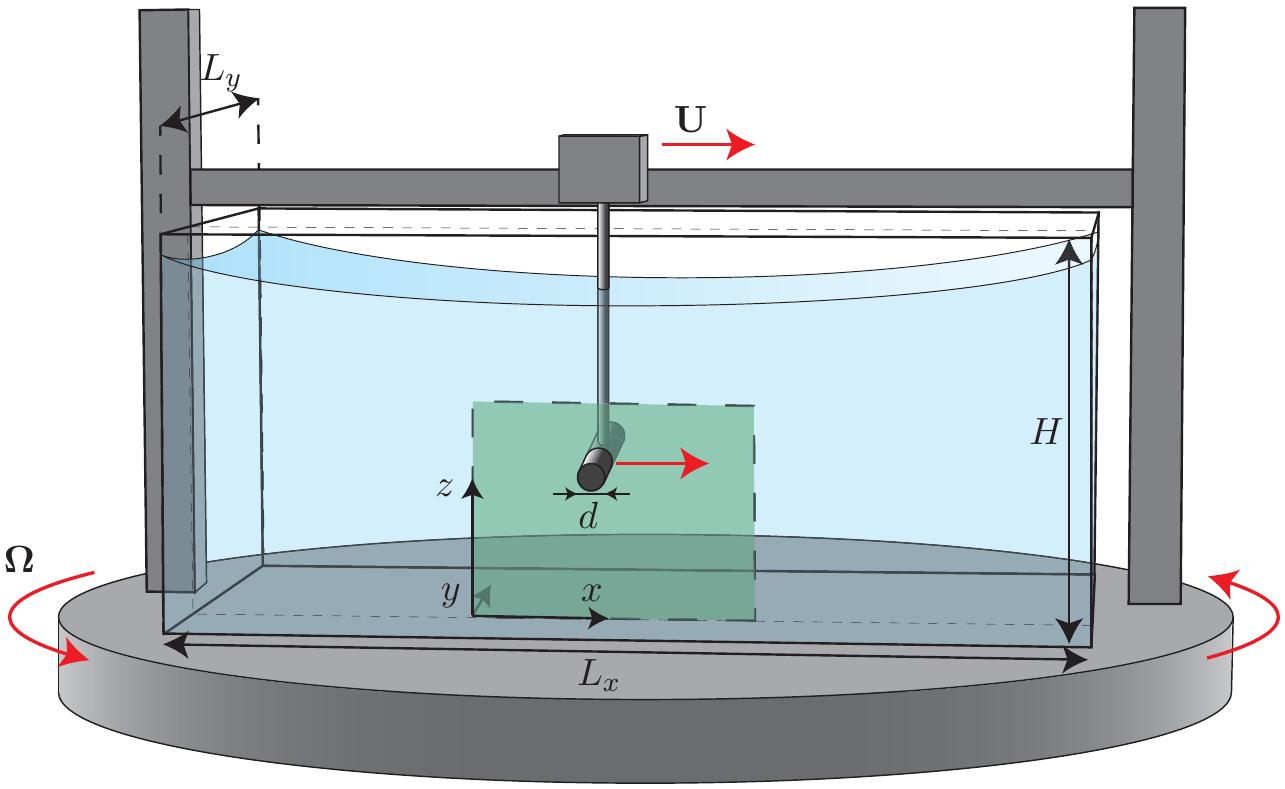}}
    \caption{Experimental setup: a motor coupled by a belt to a
    translation rail drives a horizontal cylinder, of diameter $d$ and
    length $65$~cm, at a constant velocity
    $U{\bf e}_x$ in a parallelepipedic water-filled tank mounted on a
    platform rotating at a rate $\Omega$ about $z$. PIV
    measurements are performed in the rotating frame, in a vertical
    region of $47 \times 35$~cm$^2$ (green area). $L_x = 150 $~cm,
    $L_y = 80 $~cm, $H=65$~cm.}\label{fig:setup}
\end{figure}

The experimental setup is sketched in Fig.~\ref{fig:setup}. A
horizontal cylinder, $65$~cm long in the $y$ direction, is towed
horizontally along $x$ at constant velocity $U$ between $0.6$ and
$83$~mm~s$^{-1}$ using a stepper motor coupled to a translation
rail by a belt. The translation motion, $70$~cm long, takes place
in a parallelepipedic tank, of base $L_x \times L_y = 150 \times
80$~cm$^2$ and height $65$~cm, filled with $50$~cm of water. We
have used four cylinder diameters, $d=4.1, 10.1, 20.6$ and
$40.2$~mm. The whole system is mounted on a 2~m diameter platform
rotating at a constant rate $\Omega$, in the range $2-20$~rpm,
about the vertical axis $z$. The rotation of the platform is set
at least 20 minutes before the cylinder translation to avoid
transient spin-up recirculations. The ranges of Reynolds and
Rossby numbers explored here, shown in
Fig.~\ref{fig:param_transition}, are $0.01 \lesssim {\rm Ro}
\lesssim 20$ and $2 \lesssim {\rm Re} \lesssim 3500$.

The two components $(u_x,u_z)$ of the velocity field  are measured
in the vertical plane $y_0 \simeq L_y/3$ normal to the cylinder
axis using a particle image velocimetry (PIV) system mounted in
the rotating frame ($y=0$ is the tank front face,
Fig.~\ref{fig:setup}). The fluid is seeded with 10~$\mu$m tracer
particles and illuminated by a laser sheet generated by a
corotating 140 mJ Nd:YAG pulsed laser. Images of particles are
acquired with a $2\,360 \times 1\,776$~pixels camera in a region
of interest of $47 \times 35$~cm$^2$. Each PIV acquisition
consists of $200$ to $500$ images recorded at a rate between $0.5$
and $29$~Hz depending on the amplitude of the velocity
perturbation induced by the cylinder translation.
Cross-correlation between successive images, performed over
windows of $16 \times 16$~pixels with $50\%$ overlap, produces
velocity fields sampled on a grid of $295\times 222$~vectors with
a spatial resolution of $\sim 1.6$~mm. Image acquisition starts
after 25~cm of translation, such that a steady wake regime is
reached, and the following $\sim 40$~cm of translation is
recorded.

\subsection{Steady vs. unsteady wake}\label{sec:unsteady}

\begin{figure}
    \centerline{\includegraphics[width=0.6\columnwidth]{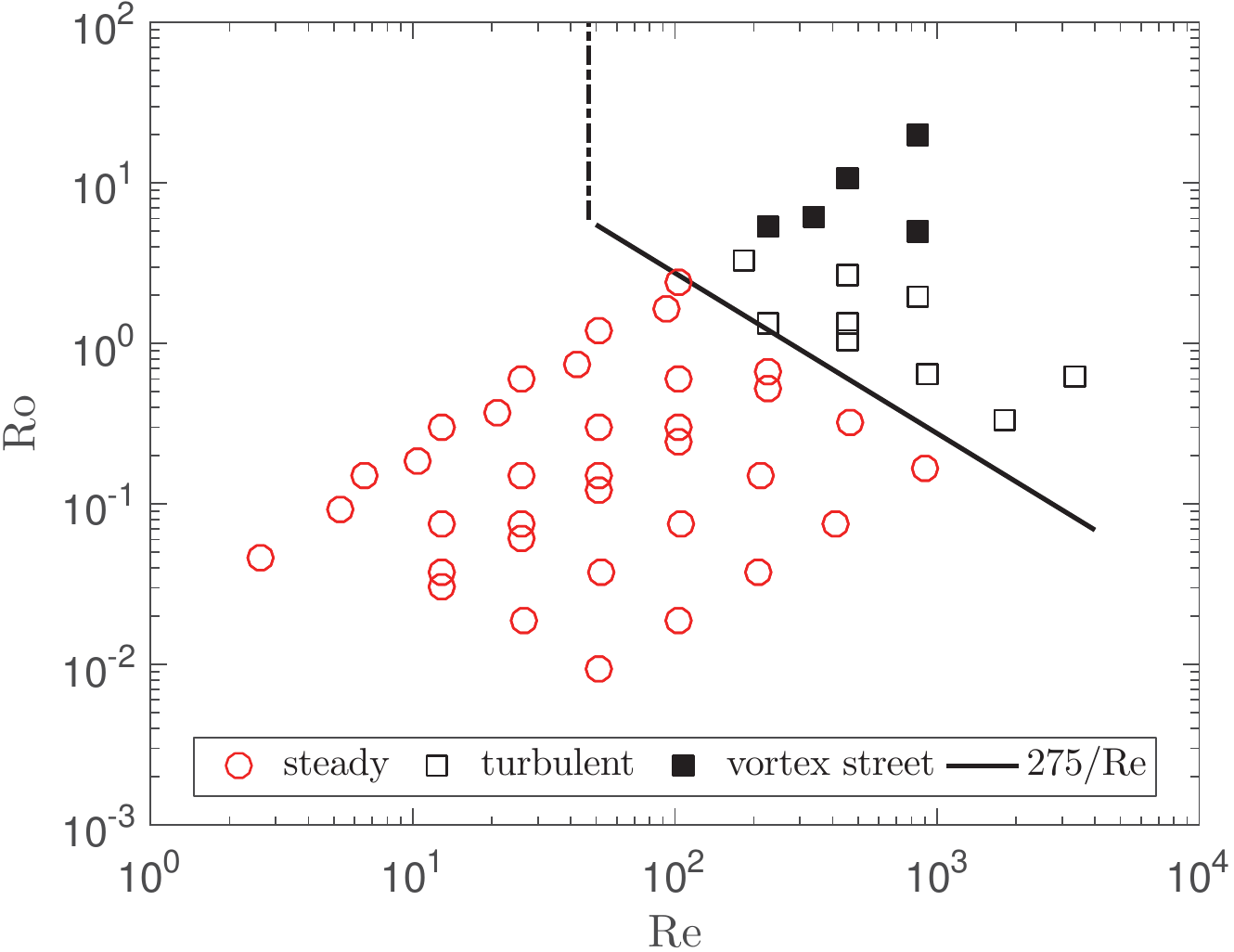}}
    \caption{Stability diagram of the cylinder wake in the (${\rm Re},
    {\rm Ro}$) plane. $\circ$: steady wake; $\blacksquare$: periodic
    vortex shedding; $\square$: turbulent wake. The continuous line
    delimits the transition to unsteadiness, ${\rm Ro}=275/{\rm Re}$.
    The vertical dashed-dotted line shows the classical instability
    threshold in a non-rotating fluid, ${\rm
    Re}=47$.}\label{fig:param_transition}
\end{figure}

We first determine the nature of the wake as a function of the two
control parameters, ${\rm Re}$ and ${\rm Ro}$. For each
experiment, we remap the velocity field in the frame moving with
the cylinder $(X=x-Ut,Z=z)$, and subtract the cylinder velocity $U
{\bf e}_x$. From these movies (see Supplemental
Material~\cite{supp_mat}), we classify the wakes in three
categories, summarized in Fig.~\ref{fig:param_transition}: steady,
unsteady with periodic vortex shedding, and turbulent.

The most remarkable effect of the global rotation is the
stabilization of the steady wake: In a non-rotating fluid (${\rm
Ro}=\infty$), the wake becomes unsteady for ${\rm Re}>{\rm
Re}_c(\infty) \simeq 47$ through the K\'{a}rm\'{a}n vortex
shedding phenomenon~\cite{Williamson1996,Guyon2015}, whereas here
steady wakes are found up to ${\rm Re} \simeq 1000$ for the
largest rotation rate. It is worth highlighting that a similar
stabilization of the wake is observed for a horizontal cylinder
translated horizontally in a linearly stratified
fluid~\cite{Boyer1989,Meunier2012a}.

An approximate Rossby number dependance of the critical Reynolds
number for unsteadiness can be inferred from our data, ${\rm Re}_c
= (275 \pm 25)/{\rm Ro}$, indicating that the stability of the
wake is governed here by the combination ${\rm Re}\,{\rm Ro}$.
This stability criterion suggests the following scenario. In a
non-rotating fluid, stability is ensured by viscous diffusion:
separation of the boundary layer and subsequent instability of the
detached layer occur when the inertial timescale, $\tau_i=d/U$,
becomes shorter than the viscous timescale, $\tau_v = d^2 / \nu$,
yielding $\tau_d / \tau_i = {\rm Re}$ as the natural control
parameter. In a rotating fluid, Ekman pumping, here at the surface
of the cylinder, provides an effective diffusion mechanism on a
shorter timescale, given by the Ekman timescale $\tau_{\rm
Ek}=d/\sqrt{\nu 2\Omega}$~\cite{GreenspanBook}. Balancing
$\tau_{\rm Ek}$ and $\tau_i$ now yields $\tau_{\rm Ek} / \tau_i =
\sqrt{\rm Re \, Ro}$ as the new control parameter, in good
agreement with our data.

\begin{figure}
\centerline{\includegraphics[width=12.5cm]{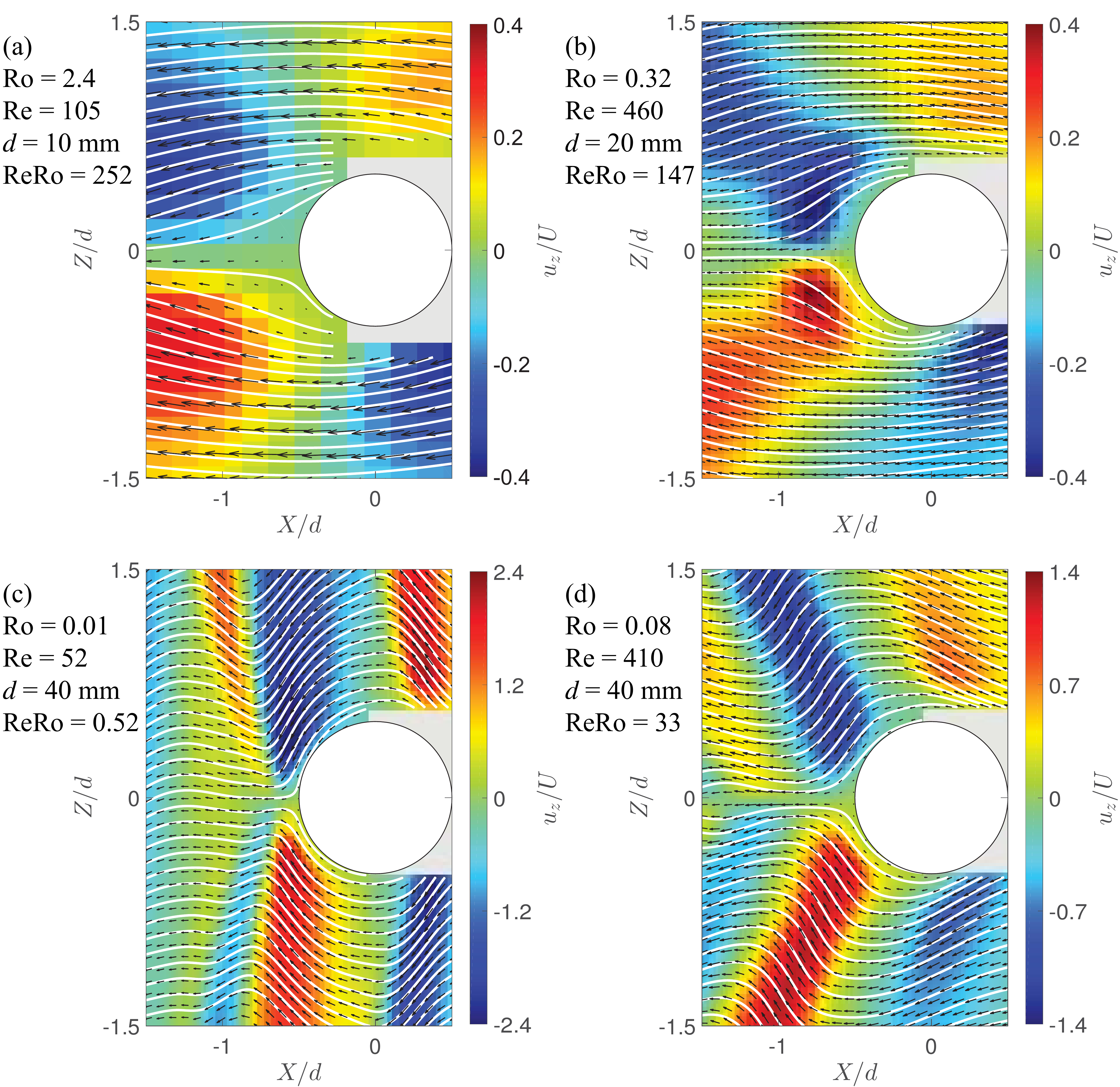}}
    \caption{Close-up view of the experimental wake of a cylinder for
    (a) ${\rm Ro}=2.4$, ${\rm Re}=105$, (b) ${\rm Ro}=0.32$, ${\rm Re}=460$, (c) ${\rm Ro}=0.01$,
    ${\rm Re}=52$ and (d) ${\rm Ro}=0.08$, ${\rm Re}=410$, for which the wake is steady.
    Each panel reports the time averaged velocity field in the
    cylinder reference frame: arrows show the in-plane velocity
    components $(u_x,u_z)$ and the colormap the vertical velocity
    component $u_z$ normalized by $U$. White lines show streamlines of
    the in-plane velocity field.}\label{fig:closeup}
\end{figure}

Close-up views of the flow near the cylinder, as shown in
Fig.~\ref{fig:closeup}, confirm the inhibition due to rotation of
the separation of the viscous boundary layers from the cylinder.
Although the PIV resolution (1.6 mm) is of the same order of
magnitude as the thickness of the Ekman boundary layer $\delta
\simeq \sqrt{\nu / 2\Omega}$, good insight into the nature of the
flow is provided by plotting the streamlines, computed here from
the in-plane velocity components. Note that the real streamlines
are helicoidal. The streamlines shown here actually correspond to
the in-plane projection of the 3D streamlines (thanks to the
invariance of our problem along the out-of-plane direction). We
find that the streamlines closely follow the back of the cylinder
up to the critical Reynolds number ${\rm Re}_c = (275 \pm 25)/{\rm
Ro}$, whereas, in a non-rotating fluid, separation occurs at ${\rm
Re}\simeq 3-4$, well before the wake instability at ${\rm
Re}_c\simeq 47$. Here, above the critical Reynolds number, the
boundary layer detaches and directly becomes unstable.

\begin{figure}
\centerline{\includegraphics[width=0.9\columnwidth]{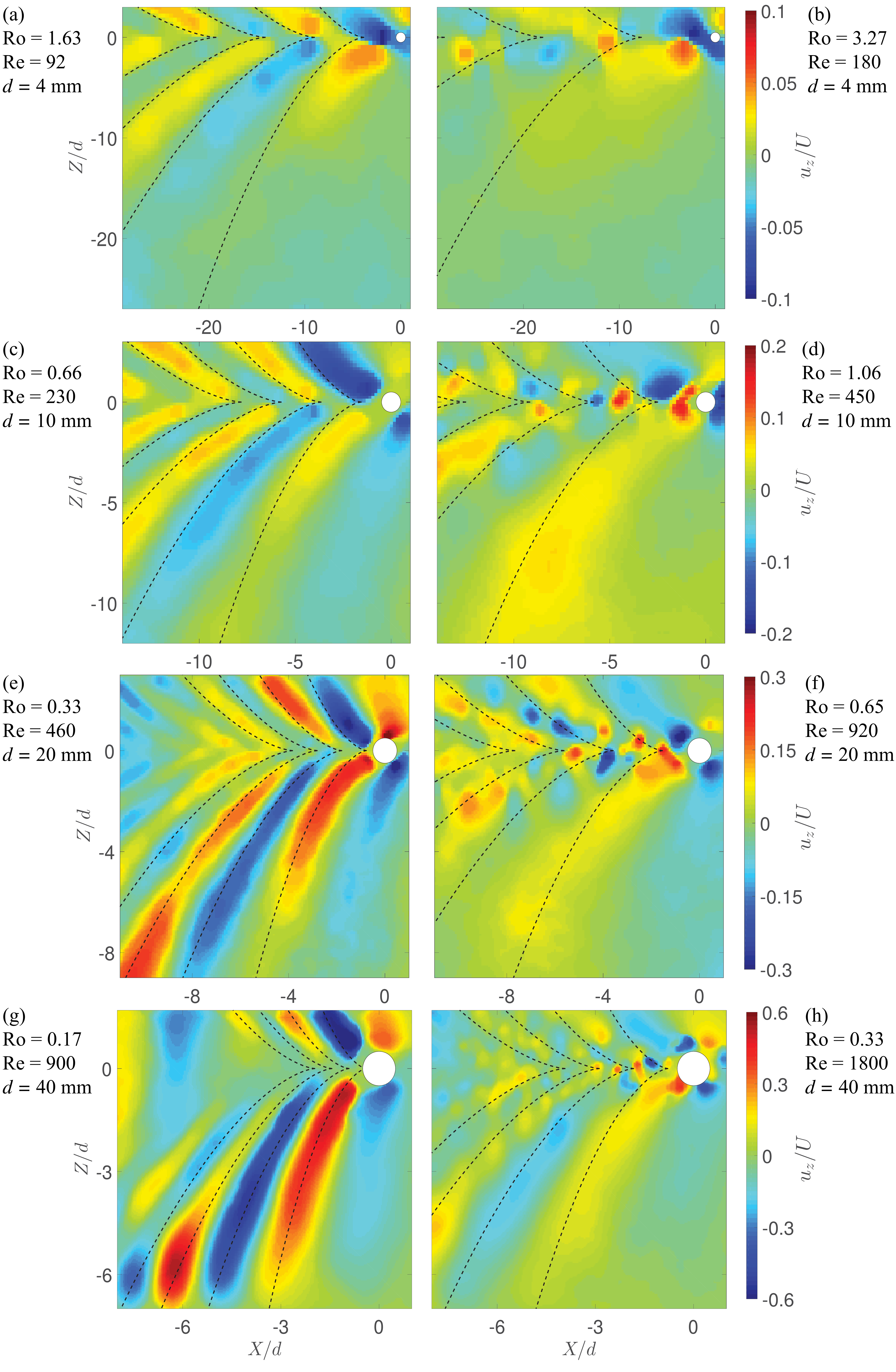}}
\caption{Instantaneous fields of the vertical velocity component
$u_z$ normalized by $U$ for several couples (${\rm Re}$, ${\rm
Ro}$) close to the onset of non-stationarity. Left column
(a,c,e,g) reports steady wakes and right column (b,d,f,h) unsteady
wakes. We show as dashed lines theoretical lines of constant phase
(\ref{eq:viswake1}-\ref{eq:viswake2}) of the far-field wake
(\ref{eq-farvx}-\ref{eq-farvz}) for $\varphi_s+\pi/2 = \pi$,
$2\pi$, $3\pi$, and $4\pi$, from right to
left.}\label{fig:champ_transition}
\end{figure}

\begin{figure}
\centerline{\includegraphics[width=0.6\columnwidth]{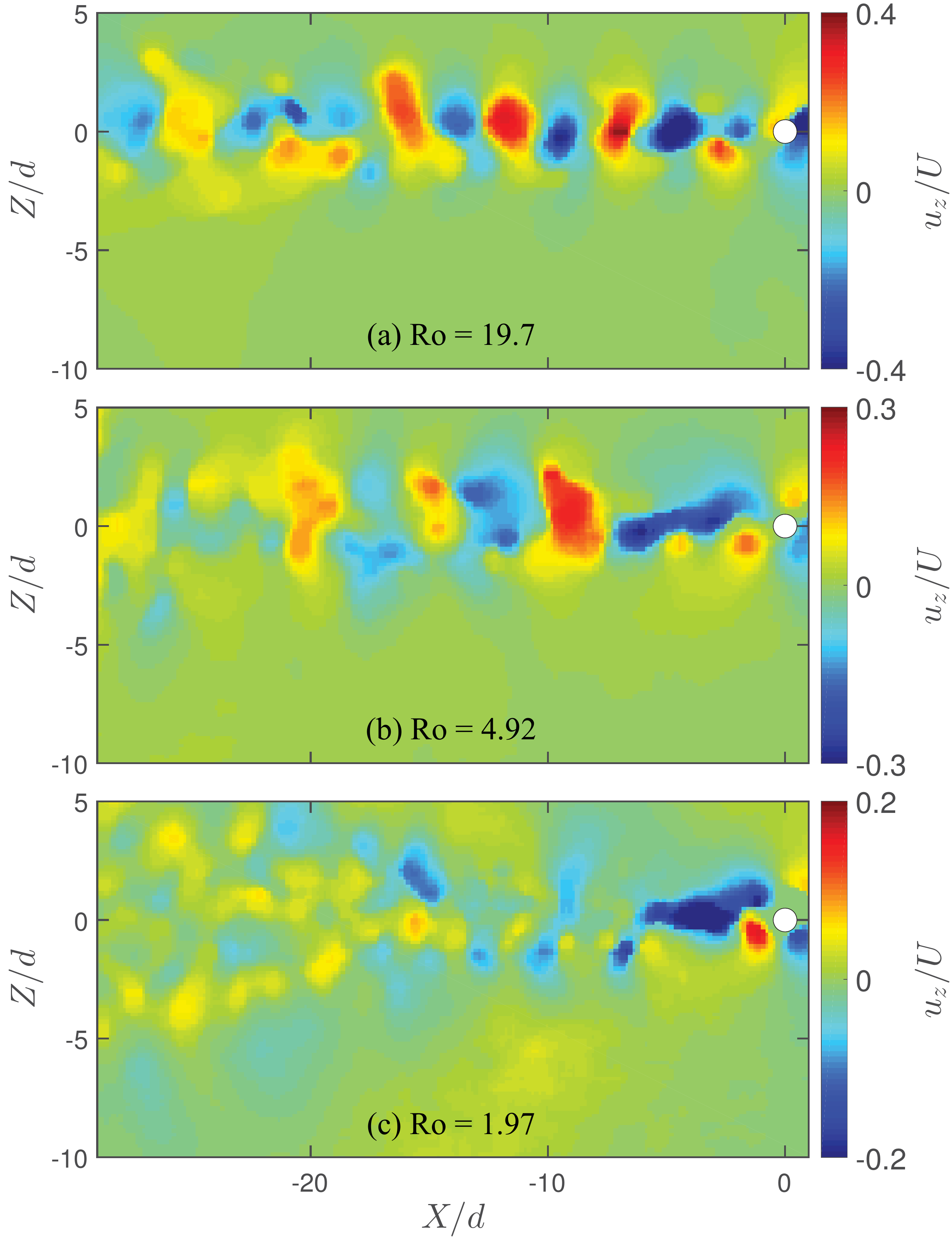}}
    \caption{Instantaneous fields of the vertical velocity component
    $u_z$ normalized by $U$ for a fixed Reynolds number ${\rm Re}=840$ and a
    decreasing Rossby number. The cylinder diameter is $d=10$~mm.}\label{fig:VK_transition}
\end{figure}

When crossing the transition line ${\rm Re}\,{\rm Ro} \simeq 275$,
the wake actually transits directly from steady to turbulent. The
nature of the wake close to the transition line ${\rm Re}\,{\rm
Ro} \simeq 275$ is illustrated in Fig.~\ref{fig:champ_transition}
for eight values of (${\rm Re}$, ${\rm Ro}$) --- four below and
four above the transition. The steady wakes, shown in
Figs.~\ref{fig:champ_transition}(a,c,e,g), have a structure in
good agreement with the predicted lines of constant phase for a
line disturbance~(\ref{eq:viswake1}-\ref{eq:viswake2}); systematic
comparisons with the model developed in Sec.~\ref{sec:theory} are
provided in Sec.~\ref{sec:steady}. The unsteady wakes in
Figs.~\ref{fig:champ_transition}(b,d,f,h) show a combination of
turbulent vortices, confined in a $\pm 20^\circ$ angular sector
centered on the wake axis, and a steady component similar to the
steady wake of inertial waves but of weak amplitude.
Interestingly, this steady component of the unsteady wakes is
still reasonably well described by the predicted phase lines for a
line disturbance. It however rapidly vanishes as ${\rm Re}\,{\rm
Ro}$ increases, as can be seen in Fig.~\ref{fig:VK_transition}.

At large Rossby number, the unsteady wake resembles a
K\'{a}rm\'{a}n vortex street but progressively becomes more and
more turbulent as ${\rm Ro}$ decreases at constant ${\rm Re}$ (see
Fig.~\ref{fig:VK_transition} for ${\rm Re}=840$ and corresponding
movies in the Supplemental Material~\cite{supp_mat}). A regular
pattern with periodic vortex shedding such as in
Fig.~\ref{fig:VK_transition}(a), typical of non-rotating
K\'{a}rm\'{a}n vortex street, is actually observed only for Rossby
numbers larger than about $5$ (filled markers in
Fig.~\ref{fig:param_transition}). For these experiments, the
shedding frequency $f$ yields Strouhal numbers $St=fd/U$ in the
range $0.15-0.20$, in good agreement with typical values found in
non-rotating fluids~\cite{Williamson1996}.

\subsection{Steady wake of inertial waves}\label{sec:steady}

\begin{figure}
    \centerline{\includegraphics[width=\textwidth]{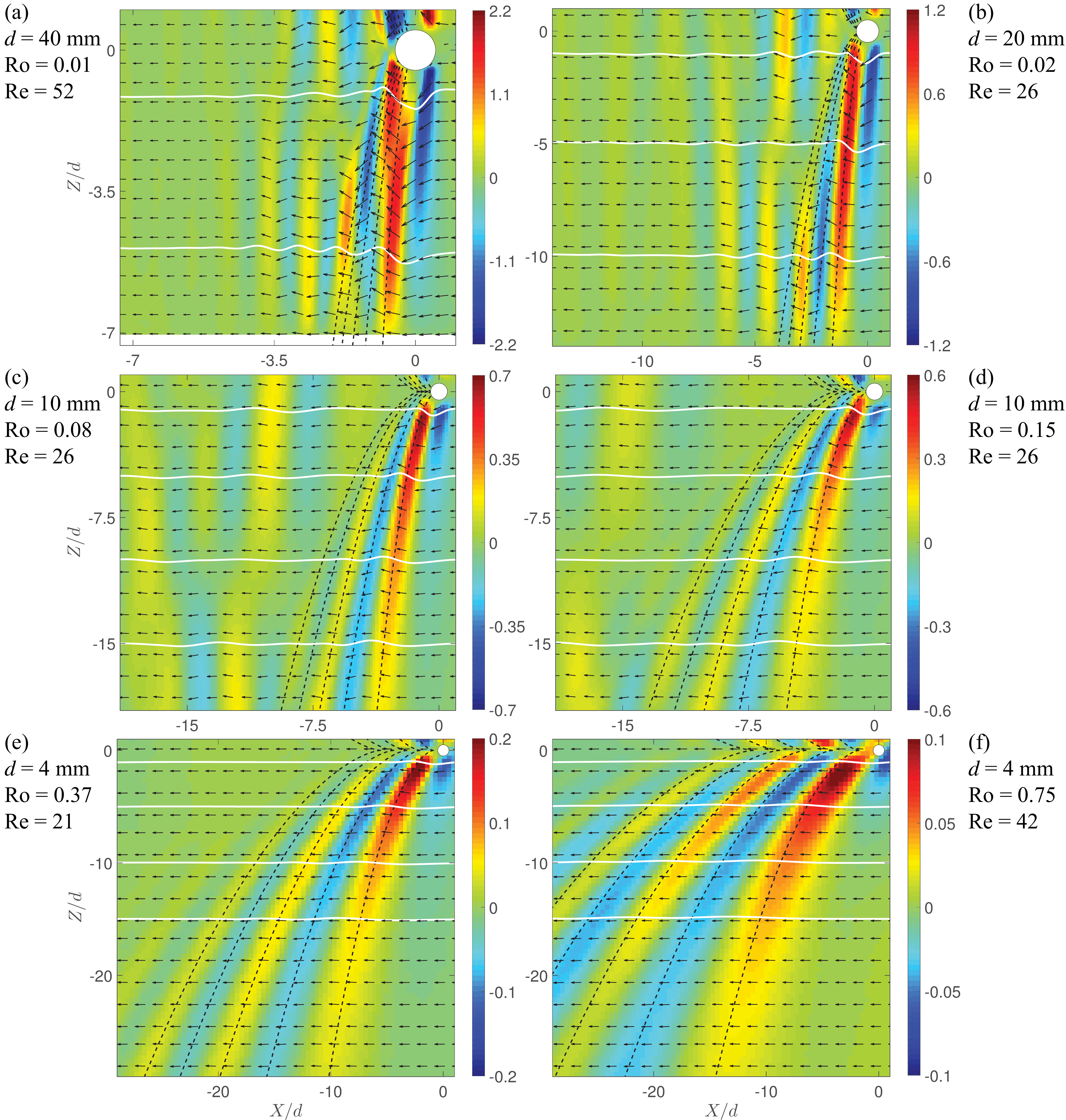}}
    \caption{Experimental wake for increasing ${\rm Ro}$ and nearly
    constant ${\rm Re}$ ($21 \leq {\rm Re} \leq 52$). Each panel shows
    the time averaged velocity field in the reference frame of the
    cylinder: arrows show the in-plane velocity components $(u_x,u_z)$
    and the colormap the vertical velocity component $u_z$ normalized
    by $U$. White lines are streamlines of the in-plane velocity
    field; assuming the flow invariance along $y$, these lines are the
    projection in the vertical plane of the real streamlines.
    Dashed lines are the predicted lines of constant
    phase~(\ref{eq:viswake1}-\ref{eq:viswake2}) of the far-field wake
    for $\varphi_s+\pi/2 = \pi$, $2\pi$, $3\pi$, $4\pi$, and
    $5\pi$.}\label{fig:exp1}
\end{figure}

In the following, we focus on the range of Reynolds and Rossby
numbers $({\rm Re}, {\rm Ro})$ for which the wake is stationary.
Snapshots of such wakes are shown in Fig.~\ref{fig:exp1}, for
increasing ${\rm Ro}$ in the range $0.01 - 0.75$, with Reynolds
numbers kept nearly constant ($21 \leq {\rm Re} \leq 52$).
Although the wakes are steady, a temporal average in the reference
frame of the cylinder is applied to filter out residual unsteady
fluid motions which are not related to the wake; these unsteady
contributions, of the order of $1$~mm~s$^{-1}$, mainly originate
from residual thermal convection and from a weak precession motion
due to the coupling of the plateform rotation with the Earth
rotation~\cite{Boisson2012}.

\begin{figure}
\centerline{\includegraphics[width=14cm]{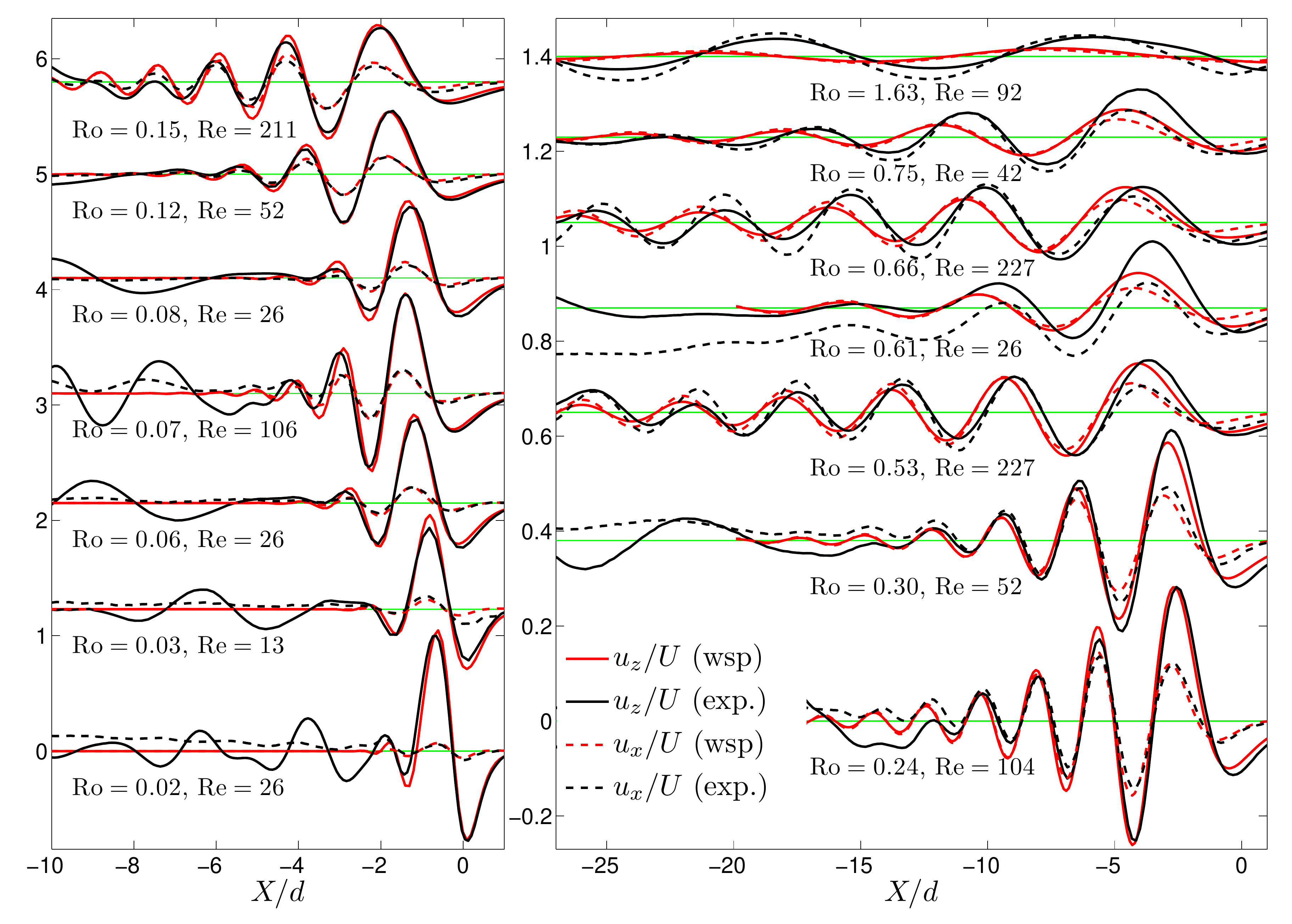}}
    \caption{Normalized axial $u_x/U$ (dotted lines) and vertical
    $u_z/U$ (full lines) velocity profiles at $Z/d=-3$ as a function
    of $X/d$ for several values of $({\rm Ro}, {\rm Re})$ for the experiments
    (black lines) and for the prediction based on the weak-streamwise-perturbation
    approximation (red lines, noted ``wsp'' in the legend).}
    \label{fig:profiles_x}
\end{figure}

\begin{figure}
    \centerline{\includegraphics[width=14cm]{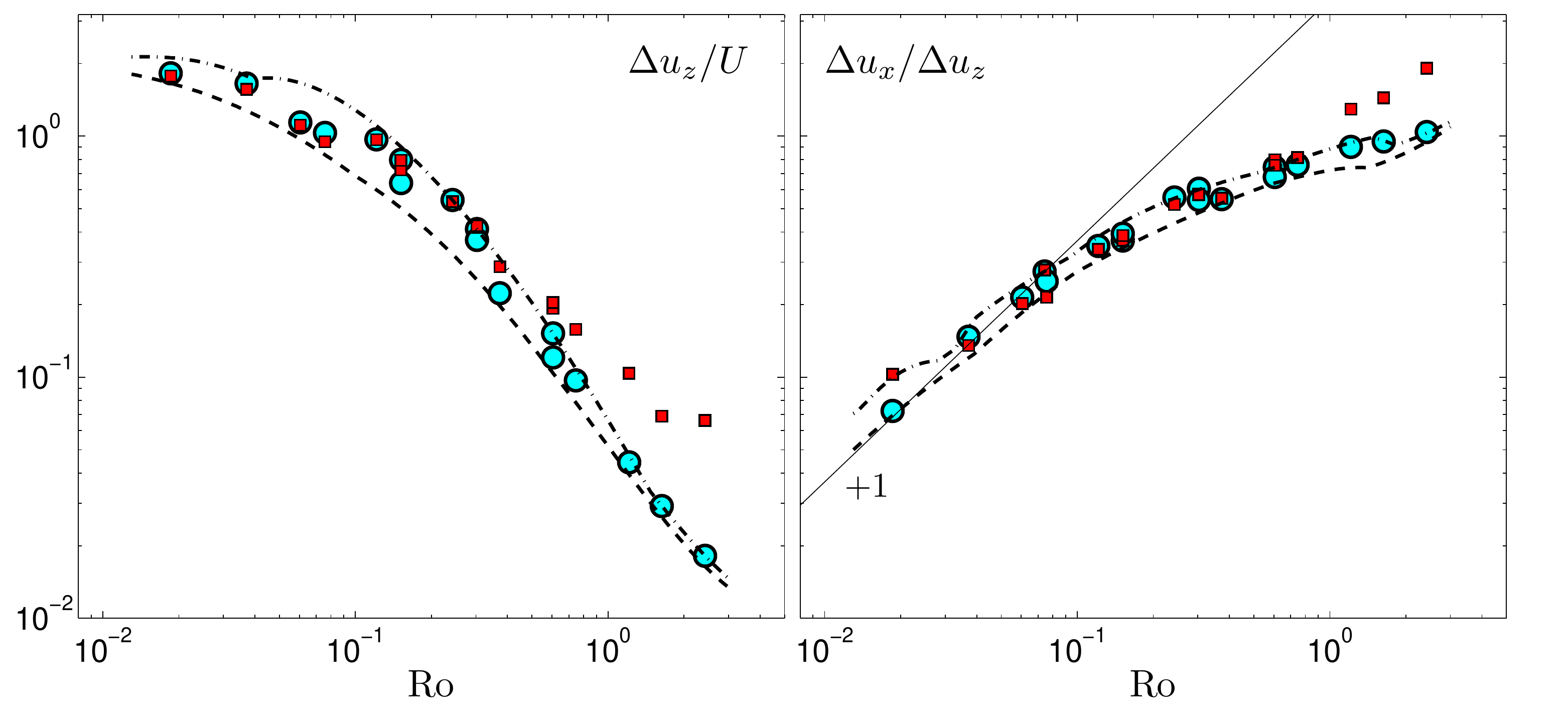}}
    \caption{Left: Amplitude $\Delta u_z/U$ of the maximum vertical
    velocity oscillation at $Z/d=-3$ as a function of ${\rm Ro}$ for
    all the experiments with $15 \lesssim {\rm Re} \lesssim 100$. For
    each data point (squares), the prediction based on the
    weak-streamwise-perturbation approximation for the same $({\rm
    Ro}, {\rm Re})$ values is reported with a circle. Right:
    corresponding ratio $\Delta u_x/\Delta u_z$ of the axial to
    vertical maximum velocity oscillation at $Z/d=-3$. The straight
    line show the expected asymptotic behavior at low ${\rm Ro}$,
    $\Delta u_x/\Delta u_z \simeq k_{x,{\rm extr}}^1 d\, Ro$. In both
    figures, the dashed and dashed-dotted lines show the theoretical
    predictions of the weak-streamwise-perturbation model for ${\rm
    Re}=15$ and ${\rm Re}=100$ respectively.}\label{fig:amp}
\end{figure}

For Rossby numbers typically larger than $0.15$
[Figs.~\ref{fig:exp1}(d--f)], the structure of the wake is in
excellent agreement with the predicted phase
lines~(\ref{eq:viswake1}-\ref{eq:viswake2}) for an infinitely
small object, shown as dashed lines. As the Rossby number is
decreased below $0.15$ [Figs.~\ref{fig:exp1}(a--c)], the wake
pattern is no longer correctly described by these lines of
constant phase: the wake becomes more vertically invariant than
predicted by the theory for a line object (note that these
snapshots contain additional wave beams that correspond to
reflections on the free surface and the bottom of the tank). This
discrepancy between the measured wake and the theoretical lines of
constant phase~(\ref{eq:viswake1}-\ref{eq:viswake2}), which is
larger as ${\rm Ro}$ is decreased, originates from the growing
influence of the size of the cylinder. This is natural since, when
${\rm Ro} = \lambda_0 / (2\pi d)$ becomes lower than $\sim 0.15$,
the characteristic wavelength along the axis of the theoretical
wake of a line object, $\lambda_0 = \pi U / \Omega$, becomes
smaller than the cylinder diameter $d$. This is also in agreement
with the results of the theoretical section~\ref{sec:theory},
which has revealed the finite size effects to become significant
below typically ${\rm Ro}\sim 0.1$. These observations illustrate
the need to consider the model~(\ref{eq:vx3}-\ref{eq:vz3})
accounting for the shape and finite size of the cylinder to
describe the experimental wake at low ${\rm Ro}$. This comparison
is further provided in Figs.~\ref{fig:profiles_x} and
\ref{fig:amp}.

This nearly vertically invariant wake observed at low ${\rm Ro}$
[Figs.~\ref{fig:exp1}(a,b)], of width of the order of the cylinder
diameter, is essentially composed of a slice of downward fluid
motion below the ``bow'' of the cylinder followed by a slice of
upward fluid motion below the ``stern'' of the cylinder. Although
this increasing vertical invariance is consistent with the
Taylor--Proudman theorem, this flow is not a Taylor column, which
is prohibited in the case of a 2D object (invariant along $y$). In
the limit of very small Rossby number, the wake stationarity
condition in the frame of the cylinder implies that low frequency
waves compose the wake in the frame of the fluid at rest. Then,
the only fluid motions allowed by the wave dispersion relation are
circular translations in vertical planes. Accordingly here, in the
frame of the fluid at rest, any fluid particle at a given $x$ must
describe, during the transit of the cylinder, one circular
translation in the vertical plane $(y,z)$, oriented by ${\bf e}_x$
(${\bf e}_x\cdot{\bf U}>0$) below the object and by $-{\bf e}_x$
above. Time symmetry of the wave dynamics then implies that the
fluid particle must come back to its initial position after the
transit of the cylinder: this explains why the upstream-downstream
non-symmetric separated wake flow, typical in non-rotating fluids,
tends to be inhibited at small Rossby number here, in good
agreement with the streamlines reported in Fig.~\ref{fig:closeup}.

In order to compare the experiments with the theory of
section~\ref{sec:theory}, i.e. Eqs.~(\ref{eq:vx3}-\ref{eq:vz3}),
we report in Fig.~\ref{fig:profiles_x} the measured and predicted
axial profiles of the normalized vertical $u_z/U$ and axial
$u_x/U$ velocity components at height $Z/d=-3$ for increasing
Rossby number and various Reynolds numbers. In the experimental
profiles at ${\rm Ro} \lesssim 0.1$, the unexpected oscillations
behind the cylinder, observed in the region where the theoretical
profiles are essentially flat, are due to the reflection of the
wake at the bottom of the water tank and at the fluid free
surface. Focusing on the distances $X$ where a non-flat profile is
predicted, we observe a quantitative agreement between the
experimental data and the model for $\rm Ro$ typically lower than
$0.30$, i.e. when significant finite size effects are present. For
larger Ro, we note an increasing wake amplitude and upstream phase
shift compared to the prediction. Such upstream phase shift was
also found by Heikes and Maxworthy~\cite{Heikes1982} for a ridge
made of a portion of a cylinder, but with a weaker wave amplitude.
They attributed this weaker wave amplitude to viscous damping,
which was not included in their model. In any case, at Rossby
numbers similar to those in Ref.~\cite{Heikes1982}, we observe a
quantitative agreement between the model and our experiments.

To provide further comparison with the model, we consider the
variation with ${\rm Ro}$ of the amplitude of the main oscillation
behind the cylinder. In Fig.~\ref{fig:amp}(a), we plot the
difference $\Delta u_z/U$ between the maximum and minimum vertical
velocity in the wake at $Z/d=-3$ for all the experiments with $15
\lesssim Re \lesssim 100$ (this range is chosen in order to limit
the dispersion due to viscosity while keeping a significant number
of data points). Experimental data are shown with square markers
and the corresponding model predictions with circles; the
predictions for the two limiting values, ${\rm Re}=15$ and ${\rm
Re}=100$, are shown with dashed and dashed-dotted lines
respectively. For ${\rm Ro} \leq 0.3$, we observe an excellent
agreement between theory and experiment, while for ${\rm Ro}>0.3$,
the measured wake amplitude becomes larger than the prediction, up
to a factor 4 for ${\rm Ro}=3$. We cannot test the evolution of
the discrepancy between the steady wake theory and the experiments
at ${\rm Ro}>3$ because it corresponds here to the largest Rossby
number at which a stable wake is observed (see
Fig.~\ref{fig:param_transition}).

We finally show in Fig.~\ref{fig:amp}(b) the ratio $\Delta
u_x/\Delta u_z$ of the axial to vertical maximum velocity
oscillations at $Z/d=-3$ for the experiments and the model as a
function of $\rm Ro$. We observe a good agreement between the
experiment and the model up to ${\rm Ro}\simeq 0.8$. Beside, one
can note that the linear behavior observed at very small $\rm Ro$
simply follows from the stationarity condition
(\ref{eq:stationnarity}), yielding $\Delta u_x/\Delta
u_z=Uk/2\Omega$, where $k$ is the local wavenumber. Indeed, at
small Rossby numbers, when finite size effects are important, the
dominant wavenumber in the wake scales as $ k \sim k_x \sim
2\pi/d$. Using the first extrema of the cylinder
spectrum~(\ref{eq:bccylq}) to estimate this dominant wavenumber,
we recover the observed linear behavior, $\Delta u_x/\Delta u_z
\simeq k_{x,{\rm extr}}^1 d\, {\rm Ro} \simeq 3.68\,{\rm Ro}$.

We finally note that, for  ${\rm Ro} \lesssim 0.30$ for which the
model describes quantitatively the observed wakes, the velocity
ratio $\Delta u_x/2U$ remains below 0.2, which provides a
justification for the approximation of weak streamwise
perturbation used in the model.

At moderate and large Rossby numbers, the choice of the model for
the translating object becomes unimportant for the global phase
pattern, which matches the line object model as shown for example
in Fig.~\ref{fig:exp1}(e-f). Nevertheless, understanding the flow
close to the object remains decisive in order to describe the wake
amplitude and phase origin (Fig.~\ref{fig:profiles_x} and
Fig.~\ref{fig:amp}). In this context, it is clear that for the
moderate Rossby numbers in the range $0.3 \lesssim {\rm Ro}
\lesssim 3$, the weak-streamwise-perturbation boundary condition
does not describe well the actual boundary condition for the wave
field, calling for a more accurate description of the viscous
boundary layer on the cylinder.

\section{Conclusion}

In this article, we study experimentally the wake produced by the
horizontal translation at constant velocity of a horizontal
cylinder in a fluid rotating about the vertical axis. For steady
wakes, we propose a  model of wake of inertial waves based on an
earlier model by Johnson~\cite{Johnson1982} that retains the
weak-streamwise-perturbation and infinite-depth approximations of
Johnson but relaxes the quasi-geostrophic approximation. We show
that this model describes the experiments quantitatively for ${\rm
Ro} \lesssim 0.3$ and for ${\rm Re}$ ranging from $1$ to $10^3$.
Our measurements confirm for the first time that this
approximation of a weak streamwise perturbation applied in a
free-slip boundary condition leads to an accurate description of
the wake even for a non-slender object, provided it is
horizontally invariant in the cross-stream direction. This result
follows from the fact that the vertical-to-horizontal velocity
ratio is imposed by the frequency in an inertial wave. Here, we
show that this low-Ro prediction applies even at moderate Rossby
number. Our experimental validation of the
weak-streamwise-perturbation model at low Ro opens the way for a
theoretical exploration of the influence on the wake of inertial
waves of the object's shape for arbitrary horizontal 2D bluff
body.

At large Rossby numbers, the structure of the wake is expected to
become independent of the size of the object, and we indeed
recover experimentally the wake predicted for a line object by
Lighthill~\cite{Lighthill1967} for ${\rm Ro}$ larger than $0.3$.
However, our measurements show that the
weak-streamwise-perturbation approximation does not anymore allow
to predict the correct amplitude and phase origin of the wake in
this regime. These results call for a better description of the
viscous boundary layers along the cylinder at moderate Rossby
number.

In this work, we also study for the first time the threshold in
Reynolds and Rossby numbers above which the wake becomes unsteady.
We find a strong stabilization by rotation of the wake, which
remains steady up to ${\rm Re}_c\sim 1000$ at ${\rm Ro}\sim 0.3$,
to be compared with the threshold for the appearance of the
K\'{a}rm\'{a}n vortex street, ${\rm Re}_c\sim 50$, in a
non-rotating fluid. Close-up views behind the cylinder indicate
that in this stabilized regime the boundary layer detachement is
inhibited by rotation. The stability criterion suggested by our
experiments, ${\rm Re} < (275 \pm 25) / {\rm Ro}$, indicates a
competition between the inertial and Ekman timescales, but no
general stability analysis for this problem is available yet.

All these results are for two-dimensional objects, invariant along
the cross-stream horizontal direction. For three-dimensional
objects, the prediction of the full velocity field of the wake of
inertial waves implies the identification of a relevant
description of the boundary layers on the object which is still an
open question in the general case. Indeed, the
weak-streamwise-perturbation approximation requires the object to
be slender and is no longer valid for 3D bluff bodies. It has been
applied to the calculation of the waves by Cheng and
Johnson~\cite{Cheng1982}, and compared with experiments involving
spherical caps and vertical pillars by Heikes and
Maxworthy~\cite{Heikes1982} who pointed out, in particular, a
lateral deflection that may be caused by the geostrophic flow
around the object. The same structure is also visible in the
numerical simulations of Mason and Sykes~\cite{Mason1981}. One
should remember this second difference with the 2D case treated in
this article: as ${\rm Ro}$ decreases, a Taylor-column
vertically-invariant flow~\cite{Hide1966,Stewartson1979} is
expected to appear and finally dominate the wake in fluid domains
of finite depth.

In geophysical configurations, the stratification of the fluid,
characterized by the Brunt--V\"ais\"al\"a frequency $N$, plays a
significant role in addition to rotation. The problem becomes that
of the horizontal flow at velocity $U$ over a topography of
streamwise size $L$, with the internal Froude number ${\rm Fr} =
U/NL$ a new parameter. Inertia--gravity waves are generated,
composed of transverse and converging waves contained by a
caustic, the streamwise orientation of the transverse and
converging waves being reversed depending on whether $N$ is
smaller or larger than $2\Omega$~\cite{Redekopp1975}. The
weak-streamwise-perturbation approximation has been applied to
this configuration by Cheng, Hefazi and Brown~\cite{Cheng1984} for
thin topographies.

Geophysical flows usually have low internal Froude numbers,
corresponding to strong stratification, in association to
moderately low Rossby numbers. When only stratification is
present, vertical motion is confined to a small layer of height
$U/N = L\,{\rm Fr}$ below the top of the topography, while the
fluid below that layer flows horizontally around the topography.
The horizontal surface separating the two layers is called the
`dividing streamline', more exactly a streamsurface. Since
vertical motion is the primary cause of internal wave motion, only
the portion of the topography protruding above the dividing
streamsurface contributes to the wave radiation, acting as a thin
cut-off obstacle to which the weak-streamwise-perturbation
approximation may be applied. This approach has first been evoked
by Newley, Pearson and Hunt~\cite{Newley1991}, Hunt \textit{et
al.}~\cite{Hunt1997} and Greenslade~\cite{Greenslade2000}, then
implemented by Hunt, Vilenski and Johnson~\cite{Hunt2006},
Voisin~\cite{Voisin2007} and Dalziel \textit{et
al.}~\cite{Dalziel2011}, the latter also presenting experiments
for hemispherical topography. As a result, the
weak-streamwise-perturbation approximation turns out to also be
applicable to three-dimensional bluff topography in a stratified
flow, so long as the stratification is strong. Generalization to a
rotating and stratified fluid depends on the deformation that
rotation may impose on the dividing streamsurface, a topic that
deserves further investigation.

\begin{acknowledgments}
We acknowledge M. Rabaud, S. Le~Diz\`es and E.~R. Johnson for
fruitful discussions, and J. Amarni, A. Aubertin, L. Auffray and
R. Pidoux for experimental help. This research was funded by
Investissements d'Avenir LabEx PALM (ANR-10-LABX-0039-PALM).
\end{acknowledgments}

\appendix
\section{Far-field expansion of weakly dissipative waves}\label{appendix1}

We look for the expansion of integrals of the form
\begin{equation}
  I(\lambda) = \int_a^b g(k)e^{i\lambda f(k)}\,dk,
\end{equation}
with $a$ and $b$ real, as $\lambda \to \infty$, in the particular
case where
\begin{equation}
  f(k) = f_0(k)+i\epsilon f_1(k),
\end{equation}
with $f_0(k)$ and $f_1(k)$ real and $\epsilon \ll 1$. Such
integrals are met when investigating the effect of weak
dissipation on the propagation of waves in the far field, as in
\cite{Johnson1982} and Sec.~\ref{sec:theory}. We assume that
$g(k)$ is regular on the interval of integration and $f_0(k)$ has
a simple stationary point $k_0$, such that $f'_0(k_0) = 0$ and
$f''_0(k_0) \neq 0$. For $\epsilon = 0$, the method of stationary
phase \cite[Sec.~2.7]{Bleistein1984} yields
\begin{equation}
  I(\lambda) \sim
  \sqrt{\frac{2\pi}{\lambda|f''_0(k_0)|}}
  g(k_0)
  \exp\left[i\lambda f_0(k_0)+i\frac{\pi}{4}\sign f''_0(k_0)\right].
  \label{eq-statint}
\end{equation}
When $\epsilon \ne 0$, to $O(\epsilon)$, the real stationary point
becomes a complex saddle point
\begin{equation}
  k_s \sim k_0-i\epsilon\frac{f'_1(k_0)}{f''_0(k_0)},
  \label{eq-steeppoint}
\end{equation}
such that $f'(k_s) = 0$, and at which
\begin{align}
  f(k_s) & \sim f_0(k_0)+i\epsilon f_1(k_0)
    \sim f(k_0),
    \label{eq-steepphase}
    \\
  f''(k_s) & \sim f''_0(k_0)+i\epsilon
    \left[
      f''_1(k_0)-f'_1(k_0)\frac{f'''_0(k_0)}{f''_0(k_0)}
    \right]
    \sim f''(k_0)-i\epsilon f'_1(k_0)\frac{f'''_0(k_0)}{f''_0(k_0)}.
    \label{eq-steepslope}
\end{align}
The steepest descent path through this point, oriented towards
increasing $\Re(k)$, makes an angle $\theta_p =
\pi/2-\arg[if''(k_s)]/2 \pmod{\pi}$ with the $\Re(k)$-axis, that
is
\begin{equation}
  \theta_p \sim \frac{\pi}{4}\sign f''_0(k_0)
  -\epsilon
  \frac{f''_1(k_0)f''_0(k_0)-f'_1(k_0)f'''_0(k_0)}{2[f''_0(k_0)]^2}.
\end{equation}
The method of steepest descent \cite[Ch.~7]{Bleistein1984} yields
\begin{multline}
  I(\lambda) \sim
    \sqrt{\frac{2\pi}{\lambda|f''_0(k_0)|}}
    g(k_0)
    \exp\left[-\epsilon\lambda f_1(k_0)\right]
    \\
  \mbox{}\times
    \exp
    \left\{
      i\lambda f_0(k_0)+i\frac{\pi}{4}\sign f''_0(k_0)
      -i\epsilon
      \frac{f''_1(k_0)f''_0(k_0)-f'_1(k_0)f'''_0(k_0)}{2[f''_0(k_0)]^2}
    \right\}.
  \label{eq-steepint}
\end{multline}
Two contributions are added to the phase $\lambda f_0(k_0)$: one
$O(\epsilon\lambda)$, affecting the amplitude of the waves, and
which, depending on $\lambda$ and $\epsilon$, may be $O(1)$ hence
must be retained; the other $O(\epsilon)$, affecting the phase of
the waves, and expected to be negligible. Accordingly, the
dominant effect of small non-zero $\epsilon$ is obtained by
evaluating the disturbed $f(k)$ at the undisturbed stationary
point $k_0$, following (\ref{eq-steepphase}).

\section{Viscous deformation of the lines of constant phase}\label{appendix2}

Quasi-geostrophy allowed Johnson~\cite{Johnson1982} to consider
viscosity at any Reynolds number~$\rm Re$. In the far field, the
steepest-descent method not only confirmed the exponential
amplitude attenuation expected from the group velocity theory, but
also revealed a viscous deformation of the lines of constant
phase, which starts to be significant ---relative to $\lambda_0$,
the characteristic wavelength of the inviscid wake--- at $\rm Re
Ro$ of order unity and makes the lines straighter near the origin.
Here, without quasi-geostrophy, an assumption of large ${\rm
Re}_{k_0} = U/\nu k_0 = {\rm Re Ro}$ had to be introduced to keep
the analysis tractable. Appendix~\ref{appendix1}, with $\epsilon =
1/{\rm Re Ro}$, then predicts the deformation to be negligible. In
the following, we consider this deformation nonetheless, in order
to assess its connection with the work of
Johnson~\cite{Johnson1982}.

\begin{figure}
\psfrag{X}[c][][1]{$X/\lambda_0$}
\psfrag{Z}[c][][1]{$Z/\lambda_0$}
\psfrag{P}[c][][1]{$\exp(-k_{z,i}Z)$} \psfrag{G}[c][][1]{$Re Ro=
1$}\psfrag{H}[c][][1]{$Re Ro= 10$}\psfrag{J}[c][][1]{$Re Ro= 100$}
    \centerline{\includegraphics[width=14cm]{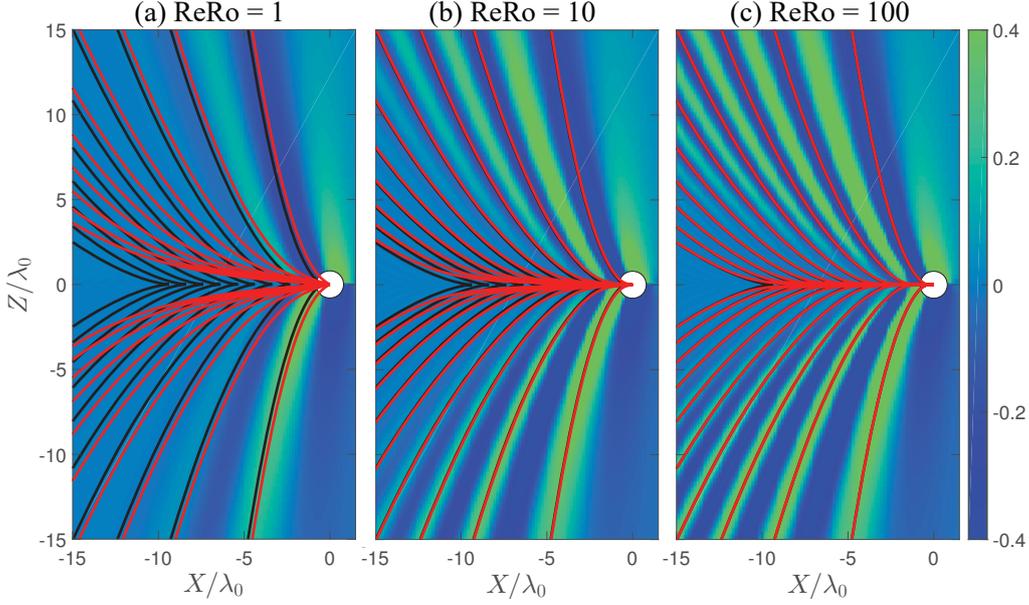}}
\caption{Lines of constant phase, computed from the viscous
model~(\ref{eq:viswake1bis}-\ref{eq:viswake2bis}) (red lines, for
$\varphi_s +\pi/2 = (2 n +1) \pi$ and for three values of ${\rm
Re}\,{\rm Ro}=1, 10, 100$) and its inviscid
limit~(\ref{eq:viswake1}-\ref{eq:viswake2}) (black lines, for
$\varphi_s +\pi/2 = (2 n +1) \pi$ and for ${\rm Re}\,{\rm
Ro}=\infty$ in which case $\varphi_s=\varphi_0-\pi/4$). The
background color shows the vertical velocity of the corresponding
prediction by the weak-streamwise-perturbation model of
Sec.~\ref{sec:modelcylinder} for a cylinder of diameter $d$ and
Rossby number ${\rm Ro}=U/2\Omega d=0.1$.}\label{fig:phaseline}
\end{figure}

The phase (\ref{eq-farphaseinv}) becomes
\begin{equation}
  \varphi_s =
  k_0|Z|\frac{\cos^2\theta}{\sin^3\theta}
  -\frac{\pi}{4}
  +\frac{1}{{\rm Re}{\rm Ro}}
  \frac{\cos\theta}{\sin^4\theta}
  \frac{6+8\cos^2\theta+\cos^4\theta}
       {4+4\cos^2\theta+\cos^4\theta},
  \label{eq-farphasevis}
\end{equation}
yielding for the lines of constant phase the parametric equation
\begin{align}
  k_0|X| & = \frac{1+\sin^2\theta}{\cos\theta}
    \left(
      \varphi_s+\frac{\pi}{4}-
      \frac{1}{{\rm Re}\,{\rm Ro}}
      \frac{\cos\theta}{\sin^4\theta}
      \frac{6+8\cos^2\theta+\cos^4\theta}
           {4+4\cos^2\theta+\cos^4\theta}
    \right),
    \label{eq:viswake1bis}
    \\
  k_0|Z| & = \frac{\sin^3\theta}{\cos^2\theta}
    \left(
      \varphi_s+\frac{\pi}{4}-
      \frac{1}{{\rm Re}\,{\rm Ro}}
      \frac{\cos\theta}{\sin^4\theta}
      \frac{6+8\cos^2\theta+\cos^4\theta}
           {4+4\cos^2\theta+\cos^4\theta}
    \right).
    \label{eq:viswake2bis}
\end{align}
The product ${\rm Re}\,{\rm Ro}$ is seen to govern the importance
of the deformation. In Fig.~\ref{fig:phaseline}, we show the lines
of constant phase $\varphi_s=2\pi n$ (for $n\in[0:10]$) for three
values of ${\rm Re}\,{\rm Ro}=1, 10, 100$ and compare them with
their inviscid counterpart $\varphi_s({\rm Re}\,{\rm
Ro}=\infty)\equiv\varphi_0-\pi/4= 2\pi n$. The background color
shows the vertical velocity of the corresponding prediction by the
weak-streamwise-perturbation model of Sec.~\ref{sec:modelcylinder}
for a cylinder of diameter $d$ and for Rossby number ${\rm
Ro}=U/2\Omega d=0.1$.

Far from the wake axis $Z=0$, as $\alpha \to
\pi/2$, the phase reduces to
\begin{equation}
  \varphi_s+\frac{\pi}{4} \sim k_0\frac{X^2}{4|Z|}
  +\frac{3}{4}\frac{|X|/|Z|}{{\rm Re}\,{\rm Ro}},
\end{equation}
and the lines of constant phase, of equation
\begin{equation}
  (4\varphi_s+\pi)|Z| \sim
  |X|\left(k_0|X|+\frac{3}{{\rm Re}\,{\rm Ro}}\right),
\end{equation}
reduce to the straightened parabolas derived by
Johnson~\cite{Johnson1982}, corrected for minor typos. Close to
the wake axis, as $\alpha \to 0$, the phase becomes
\begin{equation}
  \varphi_s+\frac{\pi}{4} \sim k_0|X|
  -\frac{3}{2}k_0|X|^{1/3}|Z|^{2/3}
  +\frac{5}{3}\frac{(|X|/|Z|)^{4/3}}{{\rm Re}\,{\rm Ro}},
\end{equation}
and the lines of constant phase, whose inviscid form exhibits
evenly spaced cusps on the axis at $k_0|X| = \varphi_s+\pi/4$, are
stretched into curves with a common corner point at the origin. As
expected, the lines of constant phase from the inviscid and
viscous models tend to coincide in the limit ${\rm Re}\,{\rm Ro}
\rightarrow \infty$. At smaller ${\rm Re}\,{\rm Ro}$, such that
the inviscid and viscous lines differ significantly, the FFT
evaluation of the vertical velocity predicted by the weak
streamwise perturbation model~(\ref{eq:vz3}) is seen in
Fig.~\ref{fig:phaseline} to better coincide with the inviscid
lines, thereby showing the viscous correction of the lines of
constant phase to be of purely rhetorical interest.

\section{Uniform far-field expansion at the wavefront}
\label{sec-uniform}

\begin{figure}
  \centerline{\includegraphics[width=\textwidth]{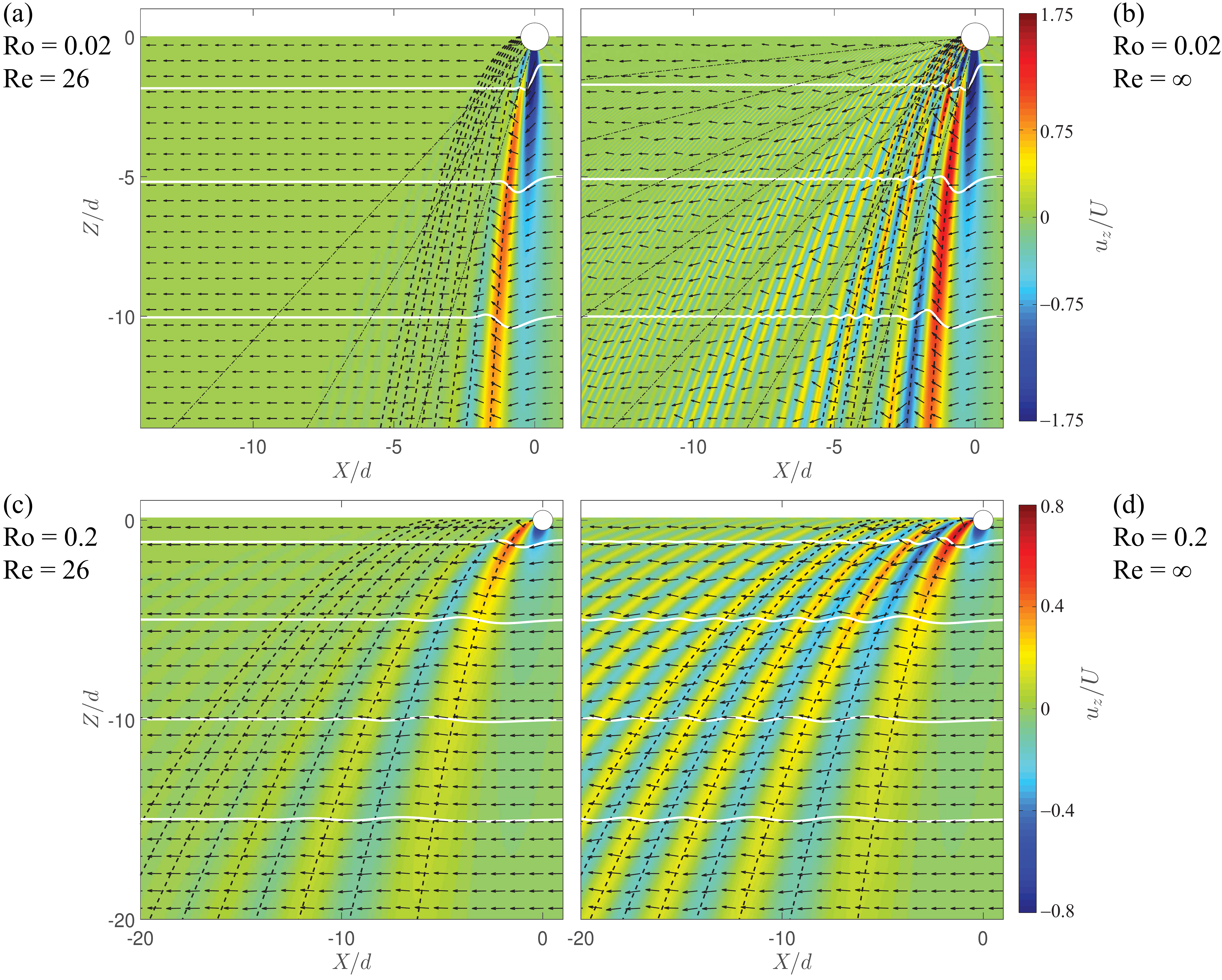}}
\caption{Wake structure predicted in the uniform far-field
approximation~(\ref{eq-uxfaruni}-\ref{eq-uzfaruni}), with the same
parameters and layout as in Figs.~\ref{fig:slender} and
\ref{fig:farfield}.} \label{fig-unifarfield}
\end{figure}

The far-field expansion~(\ref{eq-farvx}-\ref{eq-farvz}), in which
the large parameter $k_0|\mathbf{X}|$ is multiplied by
$\cos\theta$, is nonuniform at the wavefront $X = 0$ where $\theta
= \pi/2$. Mathematically, for the
integrals~(\ref{eq:vx3}-\ref{eq:vz3}), the nonuniformity is
associated with the coalescence of the stationary points $k_x =
\pm k_0\cos\theta$ with the singularity $k_x = 0$, at which the
phase of the integrand is not analytic owing to the presence of
$|k_x|$ in~(\ref{eq:kz}-\ref{eq-wavpolvisc}). Changing variables
to turn the integrals into semi-infinite ones, over $k_x > 0$, the
problem reduces to the coalescence of a stationary point with an
endpoint. Applying the analysis of Bleistein \cite{Bleistein1966},
we obtain the uniform expansion
\begin{align}
  u_x
  & = -
    \frac{\sin^{3/2}\theta\cos\theta}{\sqrt{2+\cos^2\theta}}
    \exp
    \left(
      -\frac{k_0|Z|}{\mathrm{Re}\,\mathrm{Ro}}
      \frac{\cos^3\theta}{\sin^5\theta}
    \right)
    \frac{1}{2\sqrt{\lambda_0|Z|}}
    \Re\left\{
      \widehat{q_0}(\mathbf{k}_s)
      \vphantom{\left(\sqrt{\frac{2}{\pi}\varphi_0}\right)}
    \right.
    \notag \\
  & \quad \mbox{}\times
    \left.
      \left[
        \left(
          1-(1-i)(C+iS)
          \left(\sqrt{\frac{2}{\pi}\varphi_0}\right)\sign X
        \right)
        e^{-i\varphi_s}
        -i\frac{\sign X}{\sqrt{\pi\varphi_0}}
      \right]
    \right\},
    \label{eq-uxfaruni}
    \\
  u_y
  & = -
    \frac{\sin^{3/2}\theta}{\sqrt{2+\cos^2\theta}}
    \exp
    \left(
      -\frac{k_0|Z|}{\mathrm{Re}\,\mathrm{Ro}}
      \frac{\cos^3\theta}{\sin^5\theta}
    \right)
    \frac{1}{2\sqrt{\lambda_0|Z|}}
    \Im\left\{
      \widehat{q_0}(\mathbf{k}_s)
      \vphantom{\left(\sqrt{\frac{2}{\pi}\varphi_0}\right)}
    \right.
    \notag \\
  & \quad \mbox{}\times
    \left.
      \left[
        \left(
          1-(1-i)(C+iS)
          \left(\sqrt{\frac{2}{\pi}\varphi_0}\right)\sign X
        \right)
        e^{-i\varphi_s}
        -i\frac{\sign X}{\sqrt{\pi\varphi_0}}
      \right]
    \right\},
    \label{eq-uyfaruni}
    \\
  u_z
  & =
    \frac{\sin^{5/2}\theta\sign Z}{\sqrt{2+\cos^2\theta}}
    \exp
    \left(
      -\frac{k_0|Z|}{\mathrm{Re}\,\mathrm{Ro}}
      \frac{\cos^3\theta}{\sin^5\theta}
    \right)
    \frac{1}{2\sqrt{\lambda_0|Z|}}
    \Re\left\{
      \widehat{q_0}(\mathbf{k}_s)
      \vphantom{\left(\sqrt{\frac{2}{\pi}\varphi_0}\right)}
    \right.
    \notag \\
  & \quad \mbox{}\times
    \left.
      \left[
        \left(
          1-(1-i)(C+iS)
          \left(\sqrt{\frac{2}{\pi}\varphi_0}\right)\sign X
        \right)
        e^{-i\varphi_s}
        -i\frac{\sign X}{\sqrt{\pi\varphi_0}}
      \right]
    \right\},
    \label{eq-uzfaruni}
\end{align}
with $C$ and $S$ the Fresnel functions (the term
$\sqrt{2\varphi_0/\pi}$ between parentheses is the argument of $C$
and $S$). The angles $\alpha$ and $\theta$ vary now between $0$
and $\pi$, but remain otherwise linked by~(\ref{eq-theta}). The
associated modifications to Fig.~\ref{fig:farfield} are
represented in Fig.~\ref{fig-unifarfield}.

\begin{thebibliography}{99}

\bibitem{Taylor1923}
G. I. Taylor, Experiments on the motion of solid bodies in
rotating fluids, Proc. R. Soc. A \textbf{104}, 213 (1923).

\bibitem{GreenspanBook}
H. Greenspan, {\it The Theory of Rotating Fluids} (Cambridge
University Press, Cambridge, 1968).

\bibitem{Johnson1982}
E. R. Johnson, The effects of obstacle shape and viscosity in deep
rotating flow over finite-height topography, J. Fluid Mech.
\textbf{120}, 359 (1982).

\bibitem{Hide1966}
R. Hide and A. Ibbetson, An experimental study of ``Taylor
columns'', Icarus \textbf{5}, 279 (1966).

\bibitem{Hide1968} R. Hide, A. Ibbetson, and M. J. Lighthill, On slow
transverse flow past obstacles in a rapidly rotating fluid, J.
Fluid Mech. \textbf{32}, 251 (1968).

\bibitem{Mason1981} P. J. Mason and R. I. Sykes, A numerical
study of rapidly rotating flow over surface-mounted obstacles, J.
Fluid Mech. \textbf{111}, 175 (1981).

\bibitem{Stewartson1979}
K. Stewartson and H. K. Cheng, On the structure of inertial waves
produced by an obstacle in a deep, rotating container, J. Fluid
Mech. {\bf 91}, 415 (1979).

\bibitem{Lighthill1967}
M. J. Lighthill, On waves generated in dispersive systems by
travelling forcing effects, with applications to the dynamics of
rotating fluids, J. Fluid Mech. \textbf{27}, 725 (1967).

\bibitem{Redekopp1975} L. G. Redekopp, Wave patterns generated
by disturbances travelling horizontally in rotating stratified
fluids, Geophys. Fluid Dyn. \textbf{6}, 289 (1975).

\bibitem{Peat1976}
K. S. Peat and T. N. Stevenson, The phase configuration of waves
around a body moving in a rotating stratified fluid, J. Fluid
Mech. \textbf{75}, 647 (1976).

\bibitem{Cheng1982} H. K. Cheng and E. R. Johnson, Inertial
waves above an obstacle in an unbounded, rapidly rotating fluid,
Proc. R. Soc. A \textbf{383}, 71 (1982).

\bibitem{Heikes1982}
K. E. Heikes and T. Maxworthy, Observations of inertial waves in a
homogeneous rotating fluid, J. Fluid Mech. \textbf{125}, 319
(1982).

\bibitem{Lighthill1978}
J. Lighthill, \textit{Waves in Fluids} (Cambridge University
Press, Cambridge, 1978).

\bibitem{Darrigol2005}
O. Darrigol, \textit{Words of Flow: A History of Hydrodynamics
from the Bernoullis to Prandtl} (Oxford University Press, Oxford,
2005).

\bibitem{Miles1971} J. W. Miles, Internal waves generated by a
horizontally moving source, Geophys. Fluid Dyn. \textbf{2}, 63
(1971).

\bibitem{Janowitz1984} G. S. Janowitz, Lee waves in
three-dimensional stratified flow, J. Fluid Mech. \textbf{148}, 97
(1984).

\bibitem{Voisin2007} B. Voisin, Lee waves from a sphere in a
stratified flow, J. Fluid Mech. \textbf{574}, 273 (2007).

\bibitem{Waleffe1993}
F. Waleffe, Inertial transfers in the helical decomposition, Phys.
Fluids A \textbf{5}, 677 (1993).

\bibitem{Bordes2012}
G. Bordes, F. Moisy, T. Dauxois, and P.-P. Cortet, Experimental
evidence of a triadic resonance of plane inertial waves in a
rotating fluid, Phys. Fluids \textbf{24}, 014105 (2012)

\bibitem{Kelvin1887}
W. Thomson (Lord Kelvin), On ship waves, Proc. Inst. Mech. Engrs
\textbf{38}, 409 (1887).

\bibitem{Havelock1908}
T. H. Havelock, The propagation of groups of waves in dispersive
media, with application to waves on water produced by a travelling
disturbance, Proc. R. Soc. A \textbf{81}, 398 (1908).

\bibitem{Rabaud2013}
M. Rabaud and F. Moisy, Ship wakes: Kelvin or Mach angle?, Phys.
Rev. Lett. \textbf{110}, 214503 (2013).

\bibitem{Darmon2014}
A. Darmon, M. Benzaquen, and E. Rapha\"el, Kelvin wake pattern at
large Froude numbers, J. Fluid Mech. \textbf{738}, R3 (2014).

\bibitem{Mowbray1967} D. E. Mowbray and B. S. H. Rarity, The
internal wave pattern produced by a sphere moving vertically in a
density stratified liquid, J. Fluid Mech. \textbf{30}, 489 (1967).

\bibitem{Stevenson1983} T. N. Stevenson, T. J. Woodhead, and D.
Kanellopulos, Viscous effects in some internal waves, Appl. Sci.
Res. \textbf{40}, 185 (1983).

\bibitem{Gartner1986} U. G\"artner, U. Wernekinck, and W.
Merzkirch, Velocity measurements in the field of an internal
gravity wave by means of speckle photography, Exp. Fluids
\textbf{4}, 283 (1986).

\bibitem{Torres2000} C. R. Torres, H. Hanazaki, J. Ochoa, J.
Castillo, and M. Van Woert, Flow past a sphere moving vertically
in a stratified diffusive fluid, J. Fluid Mech. \textbf{417}, 211
(2000).

\bibitem{Okino2017} S. Okino, S. Akiyama, and H. Hanazaki,
Velocity distribution around a sphere descending in a linearly
stratified fluid, J. Fluid Mech. \textbf{826}, 759 (2017).

\bibitem{supp_mat}
See Supplemental Material at [URL will be inserted by publisher]
for movies of the vertical velocity component of the wake for
(${\rm Re}=920$,~${\rm Ro}=0.65$), (${\rm Re}=840$,~${\rm
Ro}=1.97$) and (${\rm Re}=840$,~${\rm Ro}=19.7$).

\bibitem{Williamson1996}
C. H. K. Williamson, Vortex dynamics in the cylinder wake, Annu.
Rev. Fluid Mech. \textbf{28}, 477 (1996).

\bibitem{Guyon2015}
E. Guyon, J.-P. Hulin, L. Petit, and C. D. Mitescu,
\textit{Physical Hydrodynamics} (Oxford University Press, Oxford,
2015).

\bibitem{Boyer1989}
D. L. Boyer, P. A. Davies, H. J. S. Fernando, and Xiuzhang Zhang,
Linearly stratified flow past a horizontal circular cylinder,
Phil. Trans. R. Soc. A \textbf{328}, 501 (1989).

\bibitem{Meunier2012a}
P. Meunier, Stratified wake of a tilted cylinder. Part 1.
Suppression of a von K\'arm\'an vortex street, J. Fluid Mech.
\textbf{699}, 174 (2012).

\bibitem{Boisson2012}
J. Boisson, D. C\'ebron, F. Moisy, and P.-P. Cortet, Earth
rotation prevents exact solid-body rotation of fluids in the
laboratory, EPL \textbf{98}, 59002 (2012).

\bibitem{Cheng1984}
H.~K. Cheng, H. Hefazi, and S.~N. Brown, Topographically generated
cyclonic disturbance and lee waves in a stratified rotating fluid,
J. Fluid Mech. \textbf{141}, 431 (1984).

\bibitem{Newley1991}
T.~M.~J. Newley, H.~J. Pearson, and J.~C.~R. Hunt, Stably
stratified rotating flow through a group of obstacles, Geophys.
Astrophys. Fluid Dyn. \textbf{58}, 147 (1991).

\bibitem{Hunt1997}
J.~C.~R. Hunt, Y. Feng, P.~F. Linden, M.~D. Greenslade, and S.~D.
Mobbs, Low-Froude-number stable flows past mountains, Il Nuovo
Cimento C \textbf{20}, 261 (1997).

\bibitem{Greenslade2000}
M.~D. Greenslade, Drag on a sphere moving horizontally in a
stratified fluid, J. Fluid Mech. \textbf{418}, 339 (2000).

\bibitem{Hunt2006}
J.~C.~R. Hunt, G.~G. Vilenski, and E.~R. Johnson, Stratified
separated flow around a mountain with an inversion layer below the
mountain top, J. Fluid Mech. \textbf{556}, 105 (2006).

\bibitem{Dalziel2011}
S.~B. Dalziel, M.~D. Patterson, C.~P. Caulfield, and S. Le~Brun,
The structure of low-Froude-number lee waves over an isolated
obstacle, J. Fluid Mech. \textbf{689}, 3 (2011).

\bibitem{Bleistein1984} N. Bleistein, \textit{Mathematical
Methods for Wave Phenomena} (Academic Press, Orlando, FL, 1984).

\bibitem{Bleistein1966}
N. Bleistein, Uniform asymptotic expansions of integrals with
stationary point near algebraic singularity, Comm. Pure Appl.
Maths \textbf{19}, 353 (1966).

\end{thebibliography}
\end{document}